\documentclass[printer]{aa} 

\usepackage{graphicx}
\usepackage{txfonts}
\usepackage{amsmath}
\def\fdeg{\hbox{$^\circ$}}

\begin{document}

   \title{Three-dimensional extinction maps: Inverting inter-calibrated extinction catalogues}


    \author{J.L. Vergely
          \inst{1}
          \and
          R. Lallement\inst{2}
  \and
          N. L. J. Cox\inst{3}
}
   \institute{ACRI-ST, 260 route du Pin Montard, 06904, Sophia Antipolis, France\\
              \email{jean-luc.vergely@acri-st.fr}
         \and
             GEPI, Observatoire de Paris, PSL University, CNRS,  5 Place Jules Janssen, 92190 Meudon, France
                     \and
ACRI-ST, Centre d'Etudes et de Recherche de Grasse (CERGA), 10 Av. Nicolas Copernic, 06130 Grasse, France\\
}
\date{Received ; accepted }

 
  \abstract
   {Three-dimensional (3D) maps of the extinction density in the Milky Way can be built through the inversion of large catalogues of distance-extinction pairs for individual target stars. Considerable progress is currently achieved in this field through the Gaia mission. Available catalogues are based on various types of photometric or spectrophotometric information and on different techniques of extinction estimations.}
   {The spatial resolution of the maps that can be achieved increases with the spatial density of the target star{\bf s}, and, consequently, with the combination of input catalogues containing different target lists. However, this requires careful inter-calibration of the catalogues. Our aim is to develop methods of inter-comparison and inter-calibration of two different extinction catalogues. }
    {The catalogue we used as reference for inter-calibration is {\bf a} spectrophotometric catalogue. It provides a more accurate extinction than a {\bf purely} photometric catalogue. In order to reduce the dimension of the problem, a principal component analysis was performed in (G, G$_{B}$, G$_{R}$,J,H,K) multi-colour space. The subspace constituted by the two first components was split into cells in which we estimated the deviations from the reference. The deviations were computed using all targets from the reference catalogue that were located at a short spatial distance of each secondary target. Corrections and filtering were deduced for each cell in the multi-colour space.}
   {We applied the inter-calibration to two very different extinction datasets:  {\bf on the one hand, extinctions} based on both spectroscopy and photometry, representing 6 million objects and serving as a reference, and, on the other hand, a catalogue of 35 million extinctions based on photometry of Gaia eDR3 and 2MASS. After calibration, the dispersion of the extinction among neighbouring points in the second catalogue is reduced, regardless of whether reference targets are present locally. Weak structures are then more apparent. The extinction of high Galactic latitude  targets is significantly more tightly correlated with the dust emission measured by Planck, a property acquired from the first catalogue. A hierarchical inversion technique was applied to the two merged inter-calibrated catalogues to produce 3D extinction density maps corresponding to different volumes and maximum spatial resolution.  The maximum resolution is 10~pc for a 3000~pc~$\times$~3000~pc~$\times$~800~pc volume around the Sun, and the maximum size of the maps is 10~kpc $\times$ 10~kpc~$\times$~800~pc for a resolution of 50~pc. The inclusion of the spectroscopic survey data increases the dynamic range of the extinction density, improves the accuracy of the maps, and allows the mapping to be extended to greater distances to better constrain the remarkable $\simeq$2.5~kpc wide dust-free region in the second quadrant in particular, which now appears as a giant oval superbubble. Maps can be downloaded or used by means of on-line tools.}
   {}


   \maketitle
%

\section{Introduction}

As repeatedly noted for the past several years, Gaia data are both massive and of unprecedented quality. They revolutionize our knowledge of the Milky Way \citep{GAIACOLL2021}. The unique catalogue of parallaxes of Gaia  adds the third dimension of space, radial distance. The combination of information about stellar positions, velocities, atmospheric parameters, and chemical abundances adds the temporal dimension, that is, the age of stars and the global history.  In parallel, massive stellar surveys from ground or space, either photometric or spectroscopic, add a wealth of information. The surveys complement Gaia data for faint objects in the visible and add further spectral bands, especially in the infrared. Gaia and massive surveys provide all ingredients entering the determination of individual extinctions for a huge quantity of stars in the Galaxy. All extinctions, and more specifically, all measured spatial gradients of extinction, may feed tomographic inversion to obtain the 3D distribution of the Galactic extinction density \citep{Chen19, Green19, Lall19, Leike20, Rezaei20, Babusiaux20, Hottier20, Lall22}.  These 3D extinction density maps allow interpolating extinctions everywhere from the 3D maps. This estimator can replace photometric or spectro-photometric determinations if these are too uncertain or not available. In this paper, we generate the 3D extinction density by applying tomography algorithms presented in \citet{Vergely01,Vergely10}, which was extended to a hierarchical approach in \citet{Lall19}. 

The distance to individual targets can be deduced from their parallax, if available, or estimated from differences between observed and modelled luminosity. The  choice of the method depends on the parallax uncertainty. In some cases, the parallax distance may also enter a Bayesian determination along with photometric data. Extinction is always deduced from the comparison between data and models. By principle, the most accurate determination of the extinction is based on the combination of high- or medium-resolution spectroscopy and photometric measurements \citep{Sanders18, Queiroz20}. The set of observed spectral features allows constraining the temperature, gravity, and metallicity, and these spectroscopic atmospheric parameters then enter the photometric analysis. Knowing them reduces the uncertainty on the extinction considerably. Observational limitations unfortunately mean that the catalogues of targets that are observed in spectroscopy are considerably smaller than photometric catalogues, although huge progress has been achieved due to multiplex instrumentation (e.g., the Apache Point Observatory Galactic Evolution Experiment (APOGEE), the Radial Velocity Experiment (RAVE), the Large Sky Area Multi-Object Fiber Spectroscopic Telescope (LAMOST), the GALactic Archaeology with HERMES (GALAH), the Sloan Extension for Galactic Understanding and Exploration (SEGUE). Conversely, purely photometric determinations of the extinction are less precise, but the order of magnitude of the number of available targets is far higher than that for spectroscopy \citep[see ][]{Anders19}. Additionally, considerable improvement is brought by including infrared photometry in combination with blueward wavelength domains.  All 3D maps quoted above make use of infrared 2MASS photometry along with data in the visible. Finally, narrow-band photometry allows more accurate extinctions \citep{Sale14}.

Ideally, information would be maximised by using distinct extinction catalogues simultaneously in a unique inversion. This has already been attempted in the past in several ways: \textit{i}) extinction catalogues from different ground-based  photometric systems \citep{Lall14}, \textit{ii}) extinctions based on photometry and using equivalent widths of diffuse interstellar bands as proxies for the extinction \citep{Capitanio17}, and \textit{iii}) extinctions from both ground-based and Gaia photometry \citep{Lall18}. However, merging catalogues may be the source of artefacts during the inversion if some systematic differences are present, as in the following simple example. We assume that in one region of the sky, a first catalogue contains nearby stars for which extinction is overestimated, while a second catalogue contains more distant stars and their extinction is correctly estimated or slightly underestimated. During the tomographic inversion, the erroneous underestimated radial gradient found in the transition region between the two altitude ranges will result in the formation of an artificial dust layer just below this area.  {\bf This shows that catalogues must be assembled very carefully}. 

There are several types of sources for differences between the catalogues. First, each type of photometric measurement has its own calibration, which is often based on the use of non-reddened standard stars. The choice of the calibration stars differs from one technique to the next. Moreover, some standard objects may be weakly extincted. This affects the inversion at a short distance.
 Second, the stellar models that are used as references vary, and the techniques of data-model comparison vary as well. Negative extinctions are allowed or prohibited, depending on the techniques. Third, the surveys may focus on specific types (e.g. red giants). The distributions of estimated extinctions and their errors, as well as the spatial distribution of the targets, reflect this choice. Merging data for sources distributed in a specific way requires an extremely precise determination of the extinctions and their errors. Finally, errors on extinctions linked to different stellar types and/or different regions may arise {\bf from various types of} errors on distances for these object types and/or these regions. If systematic effects due to brightness and/or crowding are not perfectly contained in the quoted errors, this may produce artefacts for all catalogues that are used. 

 A simple way to inter-calibrate two sources is to use the series of objects that are common to the two catalogues and compute the relation between the extinctions from the two sources. This approach was used in the studies quoted above. This way may be appropriate if the discrepancies are related to a dominant effect, such as the extinction level, and if they do not significantly depend on stellar type. In the opposite case, using this method may become the source of errors that make the merging counterproductive. Moreover, this simple method requires the existence of a large number of common targets. Our goal here is to develop and test a more accurate method of inter-calibration that is applicable to two fully disconnected datasets. The unique requirement is that a significant fraction of the targets in each catalogue are located in the same region in 3D space.
 Section~\ref{data} describes the two catalogues we used to illustrate the inter-calibration technique. Section~\ref{calibration} describes the basic principles of the method, along with the application to the two datasets. Section~\ref{validation} contains validation tests on the individual extinctions in order to assess the inter-calibration method. Section~\ref{inversion} describes the results of several inversions of the entire dataset made from the reference and the corrected catalogues. Section~\ref{discussion} contains a summary and discusses further improvements. 
 
\begin{figure}
 \centering
\includegraphics[width=0.995\linewidth]{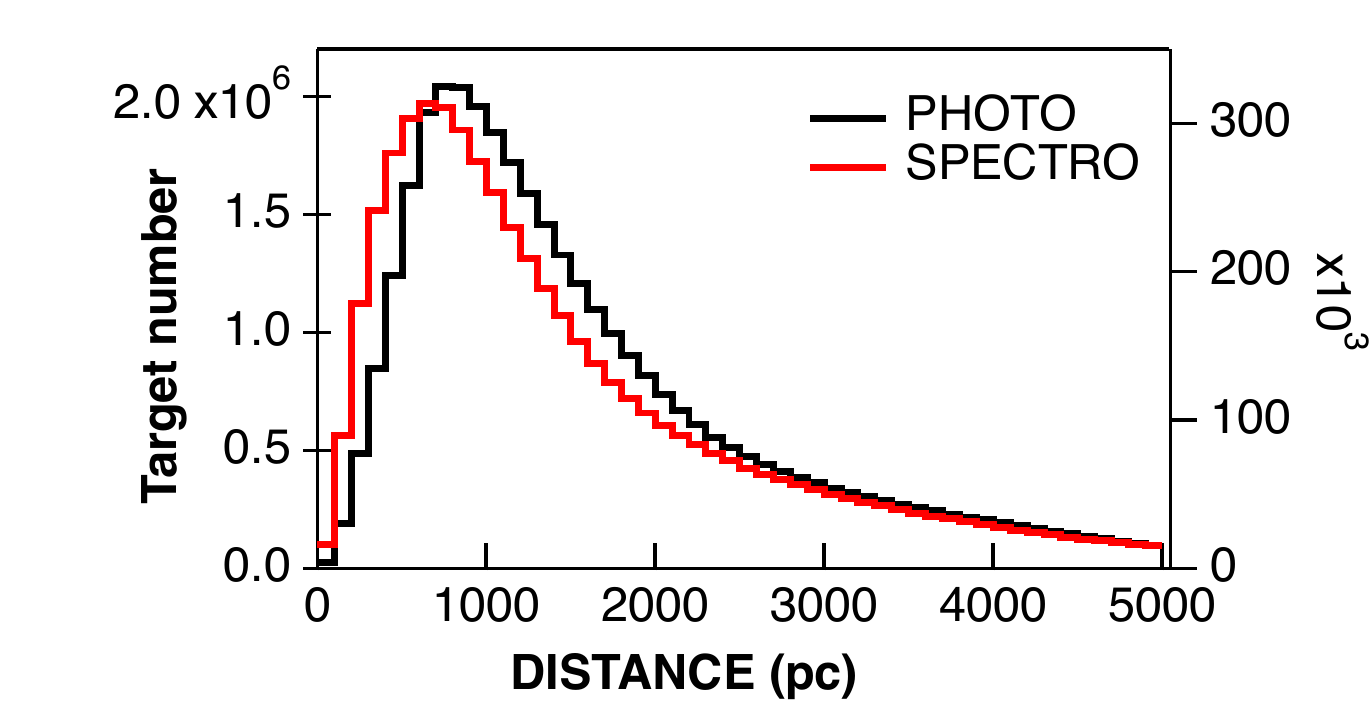}
\includegraphics[width=0.995\linewidth]{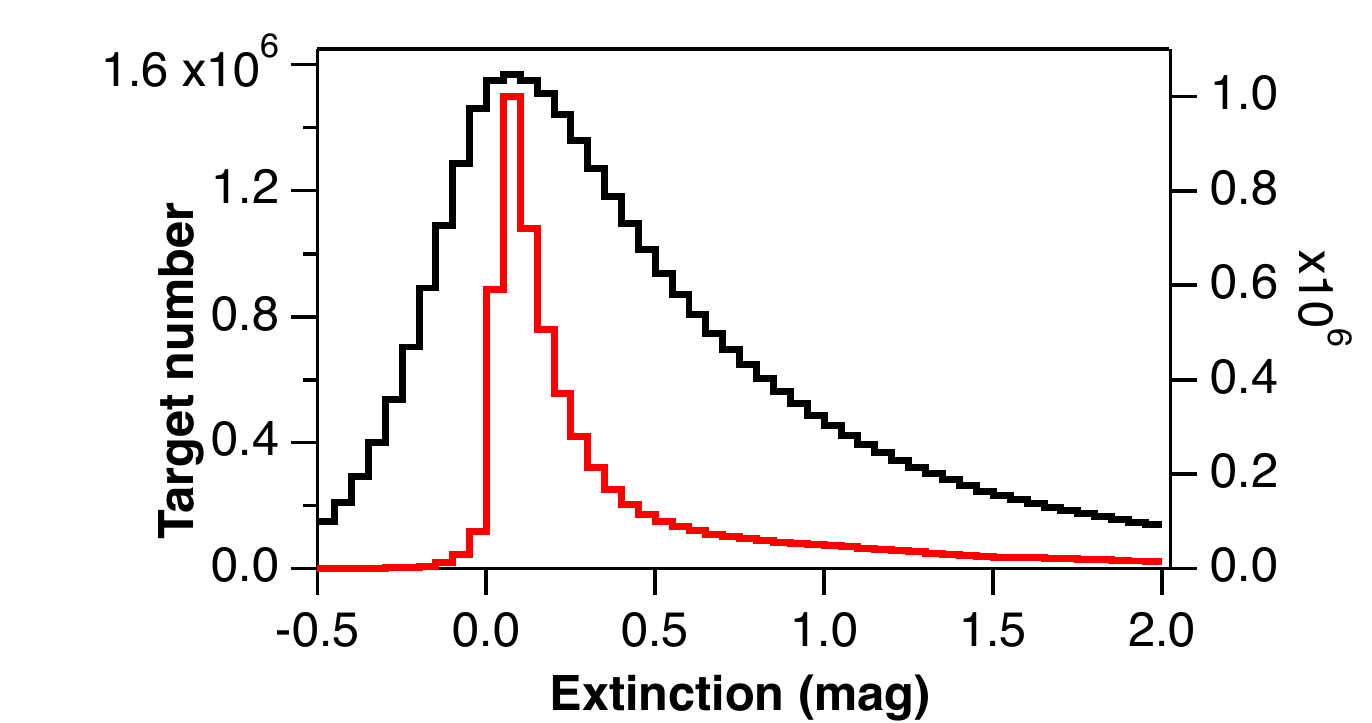}
\caption{Characteristics of the two catalogues.{\bf Top:} Histograms of target distances:  Gaia-2MASS extinction catalogue (red),  and spectro-photometric catalogue (black). The left axis corresponds to the photometric catalogue, and the right axis corresponds to the spectroscopic catalogue. {\bf Bottom:} Same as in the top panel for extinction values.}
    \label{histdistext}
\end{figure}

\section{Data}\label{data}

As already mentioned, we chose to inter-calibrate two very different types of catalogues that contain  extinctions estimated from both spectroscopy and photometry for the reference catalogue, and extinctions estimated from photometry alone for the second catalogue. In addition, the distances are estimated differently in the two catalogues. In the reference catalogue, Gaia parallaxes are used as prior values for the distances, and the final distance is determined from the parallax, the set of photometric data, and from the spectral type deduced from spectroscopy. As a result, targets with very uncertain parallaxes enter the catalogue. In the second catalogue, the accuracy of the Gaia parallax is required, and, at variance with the reference catalogue, the distance is entirely based on the parallax. Finally, extinctions are forced to be positive in the first catalogue, which is not the case in the second catalogue.

In contrast to the input data and techniques that are very different, the target distance ranges are similar in the two sources, as shown in Fig.~\ref{histdistext} (top panel). This is a necessary condition for the inter-calibration. Subsequently, extinctions are expected to be distributed in similar intervals, as shown in Fig.~\ref{histdistext} (bottom panel). The two extinction distributions peak at about the same value, but their shapes are quite different. This is caused by two facts. First, when photometry alone is used, the uncertainties are often of the same order as the estimated extinction, and the dispersion is large. Conversely, stellar  spectra add strong constraints, and  the uncertainties on the extinction are strongly reduced, which  results in a much smaller dispersion.  The second cause is the imposed positive value for the extinction in the  spectroscopic catalogue, while negative values are permitted in the second catalogue. The very small number of slightly negative values of the extinction from the spectroscopic catalogue in Fig.~\ref{histdistext} is due to the empirical inter-calibration of the two sources (see next section and the appendix).

\subsection{Catalogue based on spectroscopic surveys}\label{spectrodata}

The reference catalogue is based on two analyses by \cite{Sanders18} and \cite{Queiroz20} of ground-based spectroscopic survey data {\bf and associated} available photometric and astrometric measurements of the survey targets. The authors used the stellar parameters derived from the initial spectroscopic analysis of the surveys as input as well as Gaia DR2 parallaxes, when available, and performed a Bayesian estimate of all parameters, including extinctions, distances from astrometry and photometry, and ages.  The photometric data used in both works are from the SDSS (u; g; r; i; z), Pan-STARRS (gP; rP; iP; zP), 2MASS (J; H; Ks), APASS (B;V), WISE (W1;W2), and Gaia (G, G$_{B}$, G$_{R}$, GRVS). Our goal is to illustrate the inter-calibration of two (not three) very different datasets and the subsequent correction of one of them. We therefore preliminarily merged the two spectroscopy-based catalogues. Before merging the two sources, we performed various corrections that are described in Appendix~\ref{appendix:B}.

The analysis by \cite{Sanders18} included targets from the APOGEE, LAMOST, RAVE, SEGUE, GES, and GALAH surveys \citep{Majewski17,Deng12,Steinmetz06,DeSilva15,Gilmore12,Yanny09}. The authors used the catalogues available in 2018. Their study is based on PARSEC isochrone fitting \citep{Bressan12,Chen15, Tang14} and makes use of several prior relations and parameter estimates: mass estimates from spectroscopic parameters \citep{DasSanders19}; an elaborate Galactic model for distances, ages, and metallicities (\cite{Binney14,Queiroz18,Bovy17}); and, of importance here, a Galactic extinction prior based primarily on the \cite{Green18} maps; the \cite{Marshall06} maps at low latitudes in regions not covered by the \cite{Green18} study; and finally, the \cite{Drimmel03} model in regions not covered by any of the two previous maps. We chose to use this extinction catalogue as the reference for the calibration because this extinction prior was used. The prior enhances the quality of the extinction determination. Another reason to choose this catalogue was our particular interest in the very nearby InterStellar Medium (ISM), that is, very weak extinctions, and our previous determination of the zero point for the extinction A$_V$ from this catalogue based on nearby white dwarf (WD) absorption measurements in the UV (detailed in Appendix~\ref{appendix:A}). We selected all targets flagged {\it best} from the published catalogue for a total number of 3,318,118 stars, and we removed our estimated bias of 0.01~mag in A$_V$ from all extinctions. We did not add any other flag.

\cite{Queiroz20} derived stellar parameters, including extinctions,  for the same spectroscopic surveys except for SEGUE. Their catalogue was based on the StarHorse Bayesian analysis tool \citep{Queiroz18,Anders18}. Importantly, \cite{Queiroz20} used updated catalogues, in particular, APOGEE DR16, GALAH DR2, LAMOST DR5, GES DR3, and RAVE DR6, which resulted in a significant increase in the number of targets in comparison with \cite{Sanders18}. We excluded objects that were already present in the first catalogue from this catalogue. The authors fitted PARSEC stellar evolutionary model tracks \citep{Bressan12,Marigo17}. As priors, they used the initial mass function from \cite{Chabrier03} for all Galactic components, exponential spatial density profiles for thin and thick discs, a spherical halo, and a triaxial (ellipsoid plus spherical) bulge-bar component, broad Gaussians for the age, and mass distribution functions. They used the normalisation of each Galactic component, as well as the solar position reported by \cite{Bland-Hawthorn16}. No prior on 3D extinction was introduced. In the case of APOGEE, they used the 2D A$_V$ prior provided in the survey catalogue. 
After application of filters based on several StarHorse and Gaia flags and an empirical filtering or correction based on Gaia photometry and Planck dust emission (described in Appendix~\ref{appendix:B}), the number of additional targets from this \cite{Queiroz20} catalogue is 2,375,074 targets.

\subsection{Gaia-2MASS catalogue}\label{photodata}

Our second catalogue of data is the one described in \cite{Lall22}. It is a series of 35,463,553 extinctions derived from Gaia G, Bp, and RP and 2MASS J, H, and K photometric data on the one hand, and Gaia eDR3 parallaxes on the other hand. The targets were selected according to their accurate photometry and relative uncertainties on parallaxes better than 20~\%. The extinction was estimated based on preliminary established empirical colour-colour relations that were deduced from weakly reddened stars \citep{Ruiz-Dern18} and the most recent extinction laws \footnote{https://www.cosmos.esa.int/web/gaia/edr3-extinction-law}. The method is described in detail in this article, as is the hierarchical inversion of this homogeneous dataset, following the technique described in \cite{Capitanio17} and \cite{Lall19}. The inverted 3D extinction density map has been shown to have an increased dynamic range and reaches larger distances than the earlier map derived from Gaia DR2. We have kept the catalogue in its entirety.

\section{Inter-calibration}\label{calibration}

\subsection{Basic principle}

The inter-calibration method is based on the fact that co-spatial target stars experience the same extinction, that is, regardless of the position of these targets in the colour-colour or colour-magnitude diagrams, extinction estimates from two catalogues should produce equivalent results for stars that are neighbours in space. We chose a reference catalogue and assumed that it provided the most accurate results. We then considered all photometric bands used for the second catalogue. After correcting the different fluxes in all bands for the effects of extinction (using the extinctions given in the catalogue) and deriving the de-reddened absolute magnitudes, a principal component analysis (PCA) was implemented in (G, G$_{B}$, G$_{R}$, J, H, and K) space in order to work in a low-dimensional subspace. This subspace was then divided into cells. For each star belonging to a given cell, we compared the estimated extinction with extinctions of neighbor targets from the reference catalogue located in 3D space at distances shorter than a chosen value $D$, and we estimated for each cell the systematic differences in extinction between the two catalogues. We additionally defined some criteria to exclude peculiar data. 

The technique is not based on comparisons between extinctions of targets in common to the two catalogues, that is, it does not require any overlap. It is more general and can be applied to catalogues using very different types of targets, different types of extinction estimate, photometric and/or spectroscopic sources in different wavelength domains, or various types of absorption data.
Finally, we draw attention to the fact that this method can be used for an internal calibration of a single catalogue, that is, a search for biases associated with various stellar types. This is beyond the scope of the present work. Instead, we aim at estimating corrections or filtering data from the second catalogue to ensure  compatibility with the reference.

\subsection{Choice of catalogues}

As we stated above, our goal is to develop and test an inter-calibration method on two very different types of extinction estimates. For this reason, we selected a purely photometric determination and a hybrid determination based on spectroscopy and photometry. 
The photometric extinctions (E$_{phot}$) were derived from the three Gaia photometric bands (G, G$_{B}$, and G$_{R}$) and the three 2MASS photometric bands (J, H, and K). They were obtained by comparing the observed magnitudes with reference magnitudes in this six-dimensional space \citep{Babusiaux20}. The quality of the extinction estimator depends on the stellar type of the star and therefore on its position in this multi-dimensional space. For example, and in general, the extinction estimates of hot main-sequence stars are much more accurate than the extinction estimates of cold stars \citep{Vergely98}. In some places in the HR diagram, unresolved ambiguities can be expected to lead to biases in the extinction estimate. During the process, the location of the object in multi-colour space or colour-magnitude space is displaced, and along the displacement, more than one position with a high probability of existence and different values of the extinctions can be encountered. In this case, it is difficult to solve for the ambiguity, although it helps to know the distance.

The addition of spectroscopic data provides strong constraints on the stellar type. The results on the extinction are therefore more accurate. For the same distance and luminosity, the quoted uncertainties on the extinctions in spectroscopy-based catalogues are smaller than the quoted errors in the purely photometric determination. In order to carry out the inter-calibration, the spectroscopic extinctions (E$_{spec}$) should therefore be taken as the reference, which means that by assumption, the spectroscopic extinctions are considered to be unbiased and affected by negligible errors compared to the errors of the photometric extinctions. We therefore attempted to correct the photometric extinctions for possible biases that depend on the position in multi-colour space, so that they agreed with the spectroscopic extinction measurements. In other words, the goal of this inter-calibration is above all to work with a homogeneous set of objects and to avoid systematic shifts in extinction depending on the origin of the latter. The photometric catalogue provides monochromatic extinctions A$_{0}$ at 550~nm, while the spectroscopic catalogues estimate A$_V$, the extinction in the V band. Both quantities are very similar. We expect only weak systematic differences, and these differences are taken into account in the results of the inter-calibration.

\begin{figure}
 \includegraphics[width=0.9\linewidth]{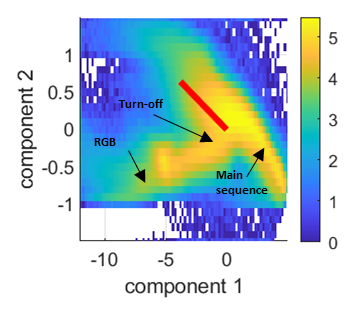}
 \includegraphics[width=0.9\linewidth]{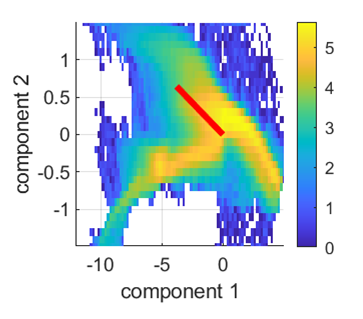} 
 \caption{Distribution of the stars of the second catalogue in C1-C2 space, where C1 and C2 are the first two principal components in the de-reddened centred reduced (Gcr, G$_B$cr, G$_R$cr, Jcr, Hcr, and Kcr) space and carry 99.9\% of the variance. The colour scale refers to the logarithm of the target number. The red line shows the average displacement of a star due to a 1~mag absorption in the visible. {\bf Top:} Before the inter-calibration. {\bf Bottom:} After the inter-calibration (see section 3.6).}
    \label{fig1}
\end{figure}

\begin{figure*}
 \includegraphics[width=0.495\linewidth]{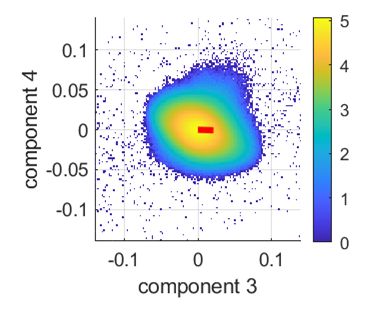}
 \includegraphics[width=0.495\linewidth]{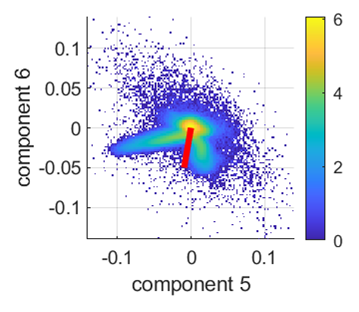}
    \caption{Same as Fig.~\ref{fig1} for C3-C4 space (left) and C5-C6 space (bottom),  which carry 0.01\% and less than 0.005\% of the variance, respectively. The colour scale refers to the logarithm of the target number.}
    \label{fig2}
\end{figure*}

\subsection{Selecting a target sub-sample for the calibration}

The inter-calibration was carried out using subsets of the photometric and spectroscopic extinction data. As the aim is to use spatially co-located stars from the two catalogues and compare their extinctions, we selected stars in a 3~kpc sphere. For the spectroscopy, we restricted the selection to stars with distance errors smaller than 25~pc and errors on the extinction estimate smaller than 0.15~mag. For the photometry, we selected objects with a parallax error smaller than 10~\%. This selection allowed us to cover the entire colour space with a significant number of objects, namely 1.2 million spectroscopic extinction values (out of a total of 5.7 million) and 29 million photometric extinction values (out of a total of 35.5 million). The purpose of this pre-selection was to calculate the inter-calibration corrections as accurately as possible. After the corrections were calculated, they were applied to the entire photometric catalogue. 
 
\begin{figure}
 \includegraphics[width=0.95\linewidth]{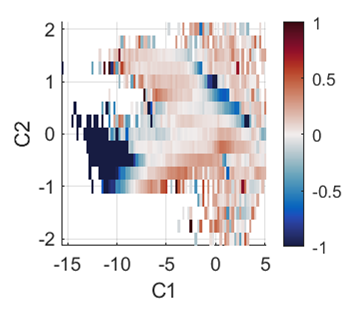}
    \caption{Mean difference E$_{phot}$-E$_{spec}$ in each cell of the de-reddened (C1,C2) plane.}
    \label{fig3}
\end{figure}

\begin{figure}
 \includegraphics[width=0.995\linewidth]{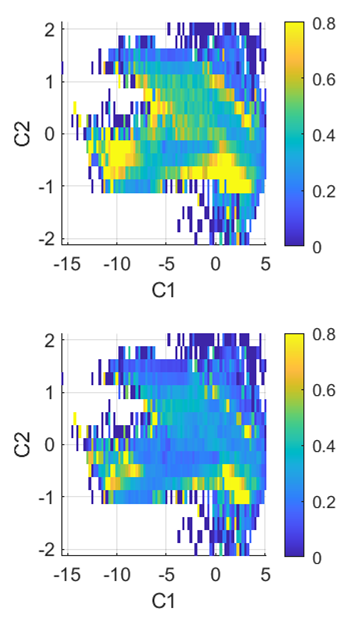}
    \caption{Standard deviation of E$_{phot}$-E$_{spec}$ in each cell of the de-reddened (C1,C2) plane. Top: Classical standard deviation. Bottom: Robust standard deviation.}
    \label{fig4}
\end{figure}

\begin{figure}
 \includegraphics[width=0.995\linewidth]{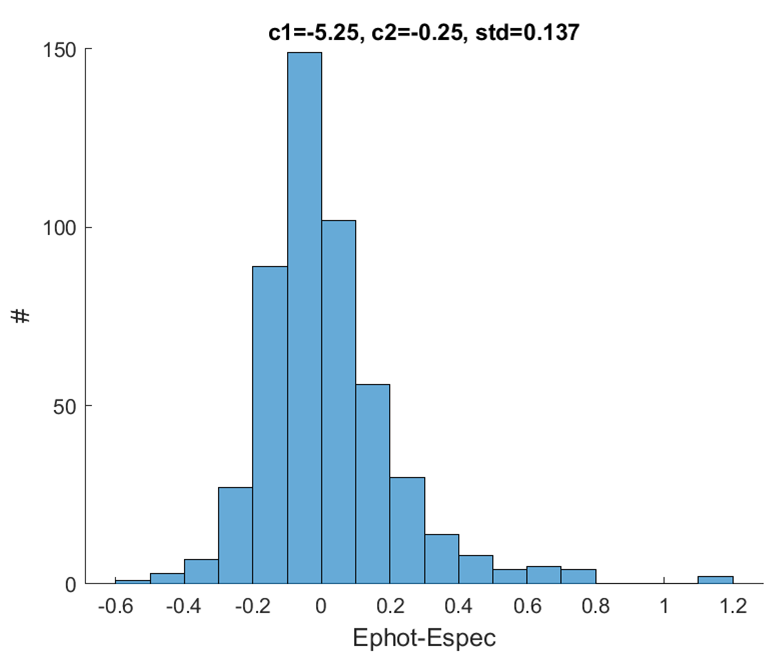}
 \includegraphics[width=0.995\linewidth]{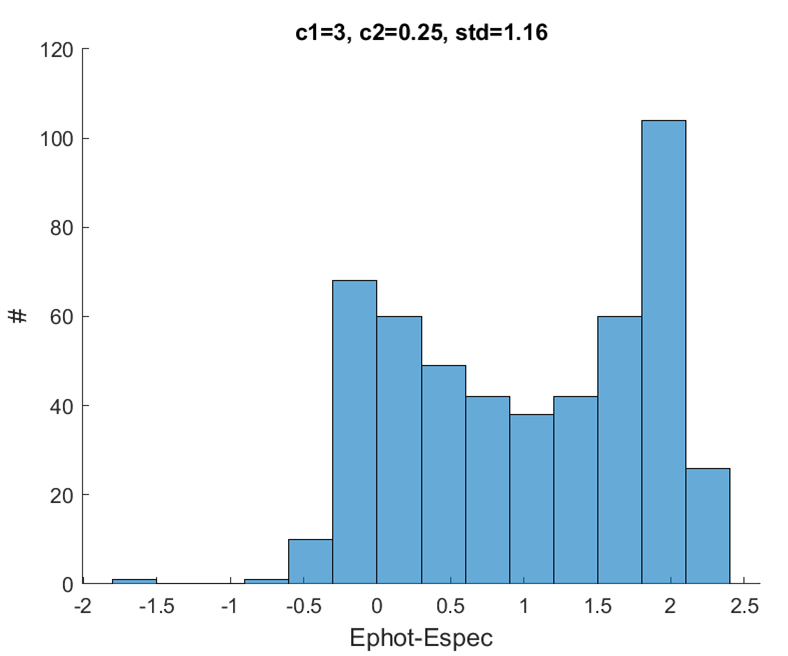}
   \caption{Two illustrative examples of E$_{phot}$-E$_{spec}$ histograms. Top: (C1,C2) plane cell with a weak standard deviation. Bottom: Same as in the top panel for a strong standard deviation.}
    \label{fig5}
\end{figure}

\subsection{De-reddening and choice of colour-colour space}

The first step was the calculation of the absolute de-reddened magnitudes of the stars from the photometric catalogue. The determination of the band-by-band extinctions was obtained by using the absorption in the visible provided by the catalogue, which was transformed to the other magnitudes (G, G$_{B}$, G$_{R}$, J, H, and K) using the extinction coefficients given by the appropriate Gaia link\footnote{https://www.cosmos.esa.int/web/gaia/edr3-extinction-law}. 
Both sets of coefficients were attempted for main-sequence stars on the one hand, and for giants and for the top of the main sequence on the other hand. {\bf These extinction coefficients account for the wide bandpasses (especially the G band) and the resulting influence of the star temperature on the extinction, and their use reduces this well-known source of uncertainties.} It has a shortcoming, however. This operation  requires an iterative calculation due to the implicit form of the equations. The iterative calculation may induce classification errors if there are problems with the convergence or/and if there are bi-modal solutions. This happened in very rare cases, and we eliminated the corresponding targets. A small number of objects was rejected at this stage (about 20,000). All other objects were de-reddened in all bands. {\bf Uncertainties on the stellar type due to the wide bandpasses are also reduced by the inter-calibration technique itself, namely the fact that we compare the extinctions of targets located in the same volume, in which all stellar types co-exist in general}.

The second step was to define the space in which the inter-calibration was to be performed. In order to normalise the magnitudes before the PCA, they were centred and reduced (G becomes Gcr, etc.): For each magnitude, the mean was subtracted, and the whole was divided by the standard deviation. We then performed a PCA, which showed that 99.9\% of the variance was carried by two components that we called C1 and C2 (see Fig.~\ref{fig1}). Components 3, 4, 5, and 6 carry relatively little information in terms of variance, as shown in Fig.~\ref{fig2}. The components C1 and C2 are linear combinations of the reduced, centred, de-reddened magnitudes Gcr, G$_B$cr, G$_R$cr, Jcr, Hcr, and Kcr:

\begin{equation}
\begin{split}
\text{C1} & = \begin{aligned}[t]
       & 0.409\; \text{Gcr} + 0.411\; \text{G}_{B}\text{cr} + 0.404\; \text{G}_{R}\text{cr} + 0.411\; \text{Jcr} \\
       & + 0.408\; \text{Hcr} + 0.407\; \text{Kcr}
       \end{aligned}
\\[1ex]
\text{C2} & = \begin{aligned}[t]
       & -0.364\; \text{Gcr} - 0.151\; \text{G}_{B}\text{cr} - 0.613\; \text{G}_R\text{cr} + 0.200\; \text{Jcr} \\
       & + 0.436\; \text{Hcr} + 0.489\; \text{Kcr.}
       \end{aligned}
\nonumber
\end{split}
\end{equation}

\begin{figure}
 \includegraphics[width=0.995\linewidth]{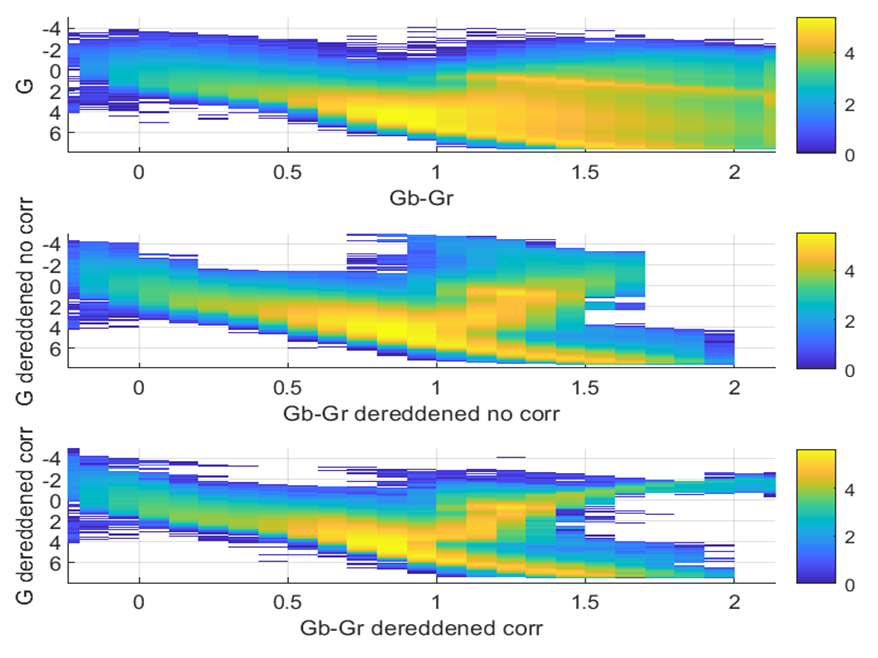}
    \caption{Distribution of stars in the (G,G$_B$,G$_R$) plane before extinction correction (top graph), after correction for initial extinction (middle), and after correction for inter-calibrated correction (bottom).}
    \label{fig7}
\end{figure}

Based on these equations, a 10~\% error in parallax amounts to an error of the order of 0.3~mag for C1 (the centred reduced normalisation is about 1.6 for all photometric bands) and close to 0 for C2 (due to sign compensations). The errors on the observed magnitudes were used as a selection criterion: less than 0.05~mag for 2MASS bands, G$_B$ and G$_R$ bands, and less than 0.02~mag on G.

The previous equation shows that C1 is a weighted sum of the absolute magnitudes, which can be seen as a kind of absolute luminosity. C2 is roughly the difference between the magnitudes (J, H, and K) and the magnitudes (G, G$_B$, and G$_R$), which gives an idea of the slope of the spectrum between the infrared and the visible. C2 therefore is a colour that is strongly correlated with the effective temperature of the star. It is interesting to note that what distinguishes one star from another in magnitude space (Gcr, G$_B$cr, G$_R$cr, Jcr, Hcr, and Kcr) happens mainly in 2D space. We work in this space for practical reasons, namely the simplification of representations and reduction of the computation times.

 

\subsection{Extinction differences}

For practical reasons, the inter-calibration was carried out in this C1-C2 2D space, which was divided into cells of size (0.25~mag $\times$ 0.25~mag). We considered that in a given cell, we can define a category of homogeneous stars that can be calibrated with the spectroscopic data. We show below that this assumption is not always verified and that it is possible to identify cells where it is not possible to inter-calibrate the data correctly. 
Each cell contains a number of objects that are spatially collocated with the objects in the spectroscopic catalogue. The objects in the spectroscopic catalogue that are within 25~pc of the stars in a given cell were selected, and we assumed that the photometric and spectroscopic extinctions are similar within the errors on average. The average difference between photometric and spectroscopic extinction gives an estimator of the correction to be made to the photometry in this given cell. It is also possible to calculate the standard deviation of the extinction difference. If this standard deviation is very large, it can give an idea of the error on the photometric extinction, knowing that the error on the spectrometric extinction is considered negligible. The standard deviation was calculated in two different ways: according to the classical method, and according to a robust method that allows decreasing the influence of extreme values ($\text{std}_\mathrm{rob}(X) = \text{median}(\text{abs}(X-\text{median}(X)))/0.6745$).

It may appear inappropriate to use reference stars located in volumes around each target instead of using those at similar distance and at small angular separation because the extinction is a line-of-sight phenomenon. However, there are complex compensating effects. The more distant the cells, the smaller the angular separations, but simultaneously, the higher the angular fluctuations of the foreground extinction. Conversely, the angle subtended by a nearby cell is wide, but the fluctuations due to the local clouds as a function of the direction are small.

Figure~\ref{fig3} shows the corrections in A$_V$ that are to be applied to the photometric extinctions in order to homogenise them with the spectroscopic extinctions. The corrections that are to be applied are smaller than 0.3~mag in most cells. However, in some specific regions of the (C1,C2) plane, especially at low values of C1, the corrections can reach 1~mag and more. {\bf As we will show below in (G, G$_{B}$-G$_{R}$) space, the effects of the corrections are significant for the branch of giants, the turn-off, and the bottom of the main sequence. In general, the HR diagram is better resolved, which suggests that the corrections are effective}.

Figure~\ref{fig4} shows the standard deviations std and std$_{rob}$ of the difference between photometric and spectroscopic extinctions in the (C1,C2) cells. When we neglect the errors in the spectroscopic extinction, then this standard deviation is a measure of the error in the photometric extinction. The representation of the differences in the form of histograms in certain areas of the plane (C1,C2) shows a bi-modal behaviour of the distributions. This means that the photometric extinction estimation is ambiguous and that there are two distinct potential solutions in this area of the colour space. When the histograms are very broad, a continuum of solutions makes the extinction estimate uncertain. Figure~\ref{fig5} shows two examples of histograms obtained in regions of the plane (C1,C2) with a weak (strong) standard deviation of the extinction differences. 

\subsection{Application of the inter-calibration}

The inter-calibration makes use of the mean value of the difference (E$_{phot}$-E$_{spec}$) and of its standard deviation in each cell.
Two resulting actions were implemented. The first action is a star-by-star correction according to the cell of the star. Figures \ref{fig1} and \ref{fig7} show the effect of the correction in the (C1, C2) and (G, G$_{B}$-G$_{R}$) planes. In both cases, the dispersion is reduced after correction. In the (G, G$_{B}$-G$_{R}$)  plane, the red giant branch is corrected in the expected direction. The second action is a filtering. If the classical standard deviation of the differences (E$_{phot}$-E$_{spec}$) was larger than 0.5~mag, then the cell is rejected, that is, the stars with photometric extinctions belonging to the cell were not used in the future tomography. We obtained a bi-modal distribution in a C1-C2 cell when there are two solutions for the reference targets that are distant in the colour-magnitude diagram because they have different extinctions. We conservatively chose to reject such cells to avoid largely erroneous estimates of the correction. The price to pay is the simultaneous rejection of some correct data points. The choice of a threshold of 0.5~  mag resulted in a loss of about 5~\% of photometric extinctions and enabled us to reject bi-modal or overly spread E$_{phot}$-E$_{spec}$ distributions.

\begin{figure}
 \centering
\includegraphics[width=0.995\linewidth]{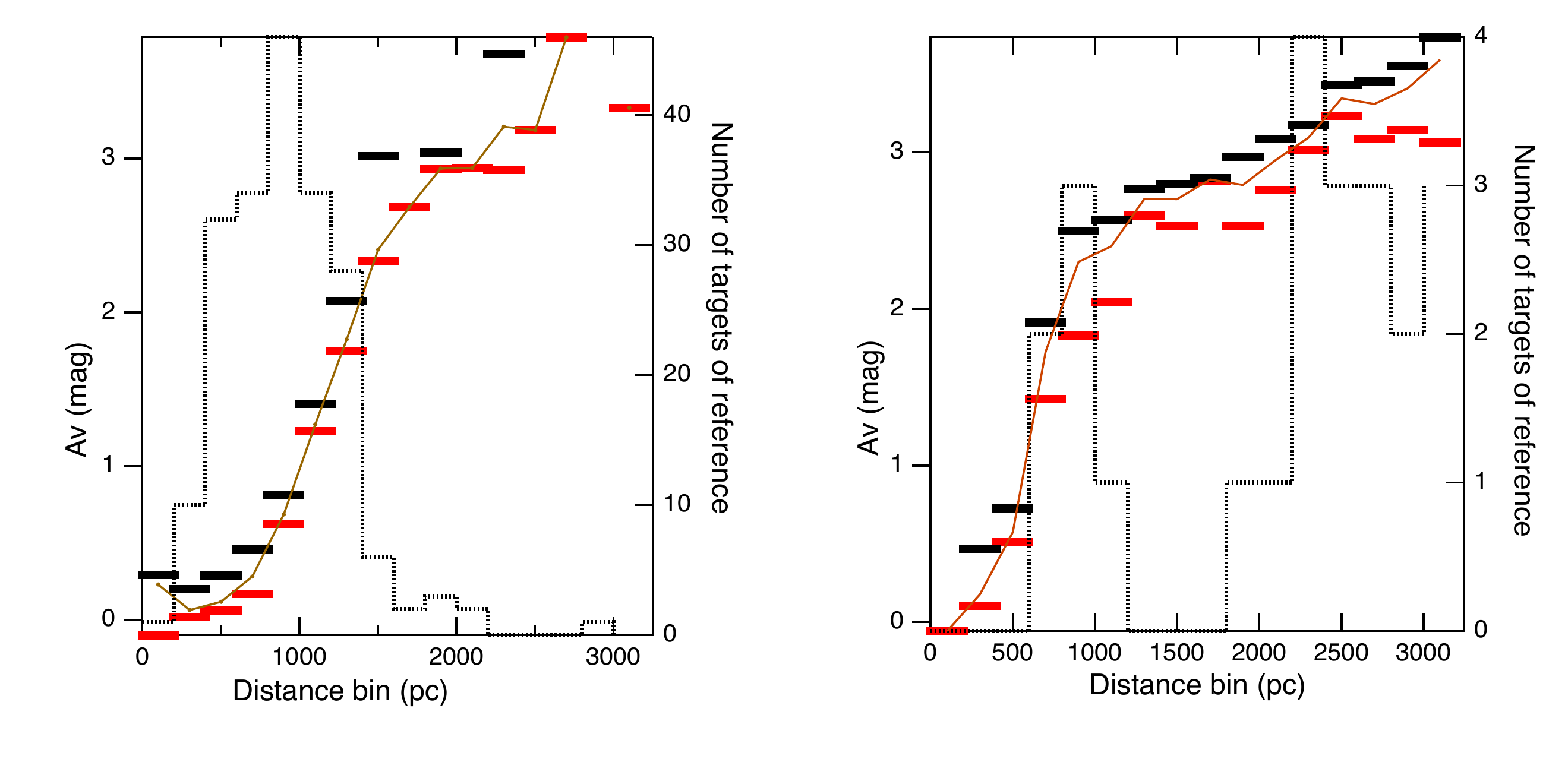}
  \caption{Illustration of the correction. {\bf Left:} Initial extinction of all targets from the photometric catalogue within 0.5\fdeg of the direction (l,b)=(60.7\fdeg, +0.2\fdeg). Data are averaged into 200 pc distance bins and separated into two groups according to the sign of the computed correction to be applied. Red shows an upward  correction (extinction increase), and black shows a downward correction. The thin black line indicates the average for all data. Black (red) points lie below (above) the average value, showing that the corrections we applied reduce the scatter. The histogram represents the number of spectroscopic extinctions in the same region of the sky and their distribution in the same distance bins. {\bf Right:} Same as left for the direction (l,b)=(76\fdeg, -2\fdeg). Significant corrections are visible in regions that are devoid of reference targets (below 600pc and between 1200 and 1800 pc.)}
    \label{senscorrection}
\end{figure}

\section{Inter-calibration validation tests}\label{validation}

There are several ways to test the effect of the correction. Some are internal, such as simple tests of changes in homogeneity and consistency of the corrected dataset. Others may use independent measurements: We performed tests of this type using the Planck optical thickness at 353 GHz \citep{PlanckXI}. 

If the reference values are accurate and the inter-calibration effectively corrects for small biases associated with the colours and the extinction level, we expect a decrease in dispersion of data points forming the line-of-sight extinction profile for any direction. This must be true also in regions of space that are devoid of targets of reference because the correction is established from the whole series of deviations measured for targets of a given type, regardless of their location. A line-of-sight extinction profile built from initial (i.e. prior to correction) distance-extinction pairs reveals a dispersion with two sources, the actual angular variability of the extinction, and the combination of uncertainties and biases. Fig.~\ref{senscorrection} displays two examples of tests illustrating the disappearance of a fraction of the dispersion. We selected from the photometric catalogue all targets located within 0.5\fdeg~ from two specific directions, here (l,b)= (60.7\fdeg, +0.2\fdeg) and (l,b)= (76\fdeg, -2\fdeg). For each direction, we separated the data into two groups according to the sign of the computed inter-calibration correction to be applied. For each group, we averaged the data into 200 pc distance bins. The figure shows that for both directions, data moving down are all above data moving up. This shows that the dispersion around the average profile, also shown in the figure, will decrease. The same trend is obtained when tests were performed at different latitudes and longitudes. The two figures also display the number of  distance-extinction pairs of stars from the catalogue of reference located in the same solid angle and the same distance bin as the photometric targets. The decrease in dispersion is not limited to regions with high densities of reference targets, but extends to regions that are devoid of these targets. This demonstrates that an effective correction is spread in space, and not only close to reference targets.

\begin{figure}
 \centering
\includegraphics[width=0.9\linewidth]{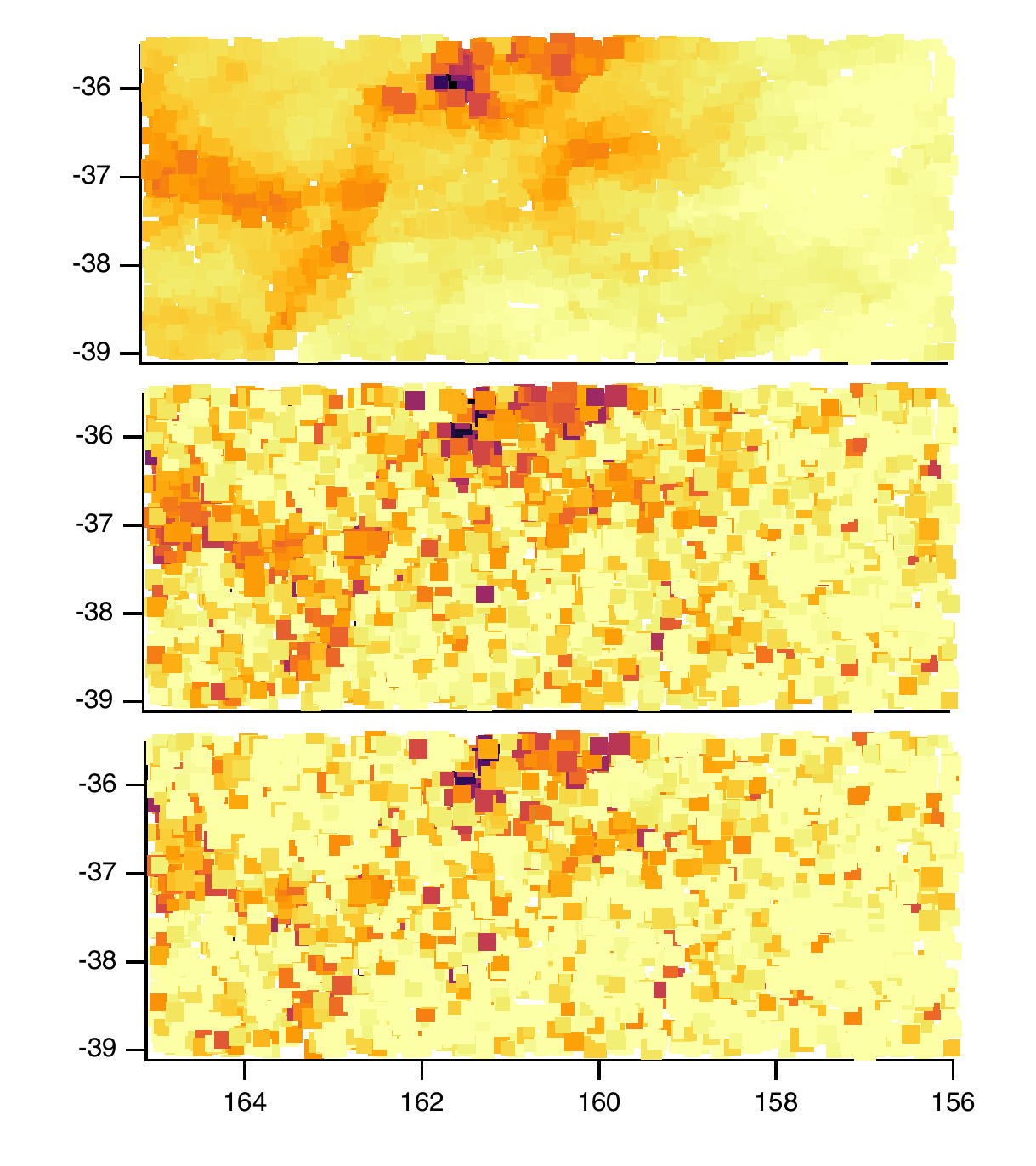}
  \caption{Illustration of the correction.  {\bf Top:} Planck $\tau$353~GHz dust optical thickness in the direction of photometric targets in a region of the sky in the low Galactic latitude part of the Taurus-Perseus area. The image is used to show weak filamentary structures. {\bf Middle:} Corrected extinction for the same targets.  {\bf Bottom:} Initial extinction for the same targets.  The colour scale is identical to the one used in the middle image. The subtle variations in the pattern in the middle image agree better with Planck structures.  Stars from the reference catalogue are not displayed in the images. }
    \label{taurus}

\end{figure}

Another test can be performed using 2D images. Fig.~\ref{taurus} displays the pre- and post-correction extinctions for targets of the second catalogue located in an area at low Galactic latitude below Perseus. There are subtle changes due to the correction in this area. To highlight these very weak changes, an image based on the same targets, this time colour-coded according to the dust optical thickness deduced from Planck data \citep{PlanckXI}, is shown for comparison. The Planck image shows that filamentary structures that are best seen with Planck appear slightly more clearly after the correction. 



\begin{figure}
 \centering
\includegraphics[width=0.995\linewidth]{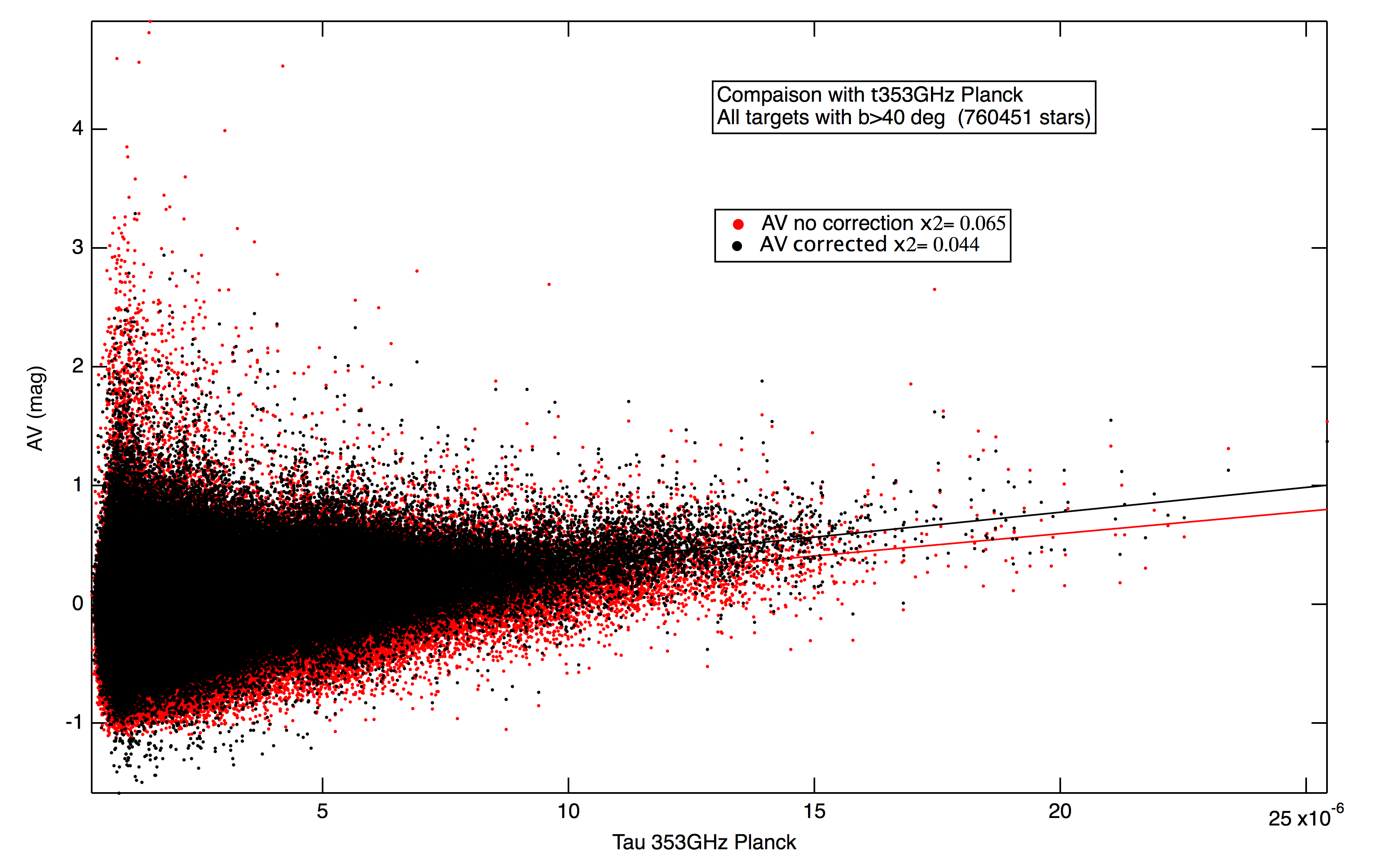}
  \caption{Illustration of the correction. Red dots show the initial extinction for all targets of the photometric catalogue above 40\fdeg Galactic latitude as a function of the optical thickness of the interstellar dust at 353GHz, as measured by Planck \citep{PlanckXI}. Black dots show the corrected extinction for the same targets. A linear fit to Planck $\tau$353 gives a significantly lower $\chi^{2}$ value for the corrected extinctions, showing that some biases have been corrected.}
    \label{highlat}

\end{figure}

Finally, a validation based on the comparison with the dust emission can be performed, provided it is restricted to high latitudes. Because the dust is located at a short distance from the Plane (within $\simeq$200 pc; see e.g. \citealt{Lall22}), the absorbing dust for most high-latitude targets is the totality of the dust seen in emission, and a tight correlation between the extinction and the dust optical thickness is expected. We used the Planck optical thickness because it provides the highest angular resolution over the sky and does not enter any extinction prior used in the inter-calibrated catalogues. 
Fig.~\ref{highlat} displays all pre- and post-correction extinctions for all targets of the second (photometric) catalogue located above $b = +40$\fdeg\  as a  function of the Planck 353~GHz optical thickness $\tau$353. For all these targets, the extinction is very weak, often of the order of the uncertainty. Linear fits to $\tau$353 are shown, and the slope is slightly higher after correction. More importantly, the dispersion around the linear fit is significantly reduced ($\chi^{2}$ decreases by 32~\%). This shows that even for these very weak extinctions, the inter-calibration brings some improvement,  and that some reliability of the spectroscopic data has been transmitted to the photometric extinctions. Part of this reliability may be due to the extinction prior used by \cite{Sanders18} for all targets or by \cite{Queiroz20} for the subset of APOGEE targets, and may be transferred to the final maps discussed in the next section.  However, as already pointed out, this transmission is not directional, but goes through the location of the targets in multi-colour space C1-C2. 



\section{Inversion of merged catalogues} \label{inversion}

We have applied the hierarchical inversion technique presented in \citep{Vergely10} and \cite{Lall19} to the merged catalogue  composed of the reference catalogue and the inter-calibrated one. This technique uses spatial correlation lengths in order to allow estimating extinction density everywhere. Several extinction density maps were computed for several maximum spatial resolutions and extents. We recall that the best achievable resolution (or minimum achievable spatial correlation length) depends on the spatial density of the target star and on distance uncertainties, that is, it decreases with the distance to the Sun and the opacity encountered in the various regions. 
For these reasons and to reduce the computational time, maps covering a large volume were computed with a larger correlation length. The series of maps correspond to volumes $\Delta$X x $\Delta$Y x $\Delta$Z of

\begin{figure*}
     \includegraphics[width=0.95\linewidth]{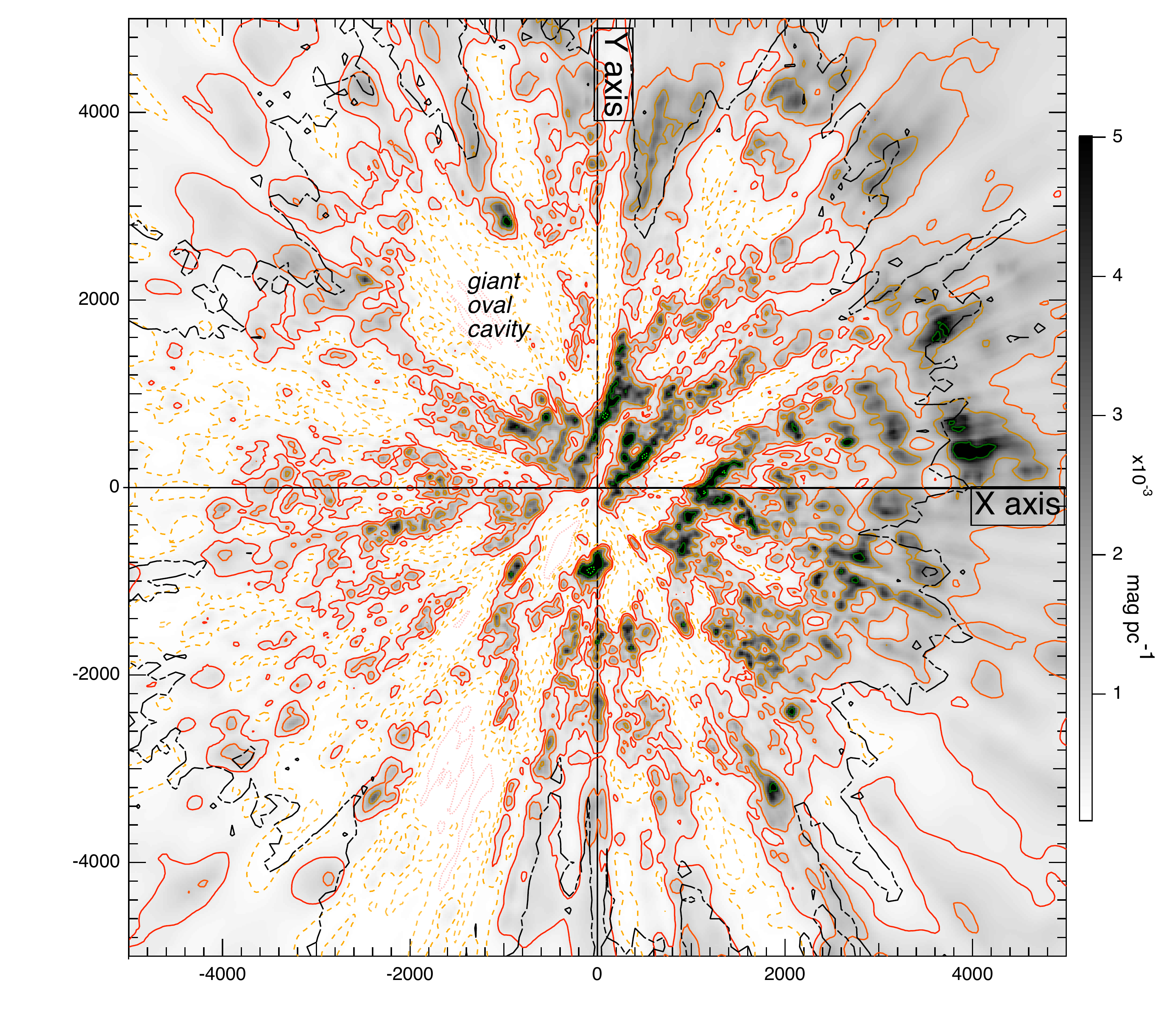}
     \centering
    \includegraphics[width=0.99\linewidth]{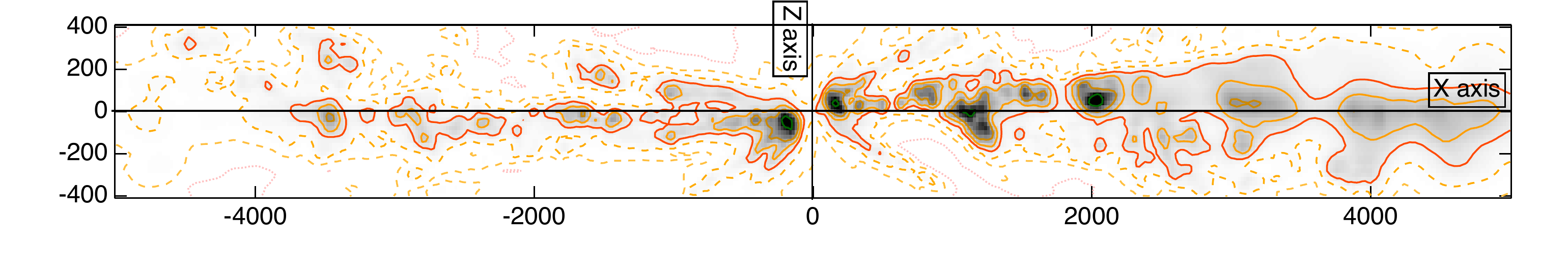}
    \includegraphics[width=0.99\linewidth]{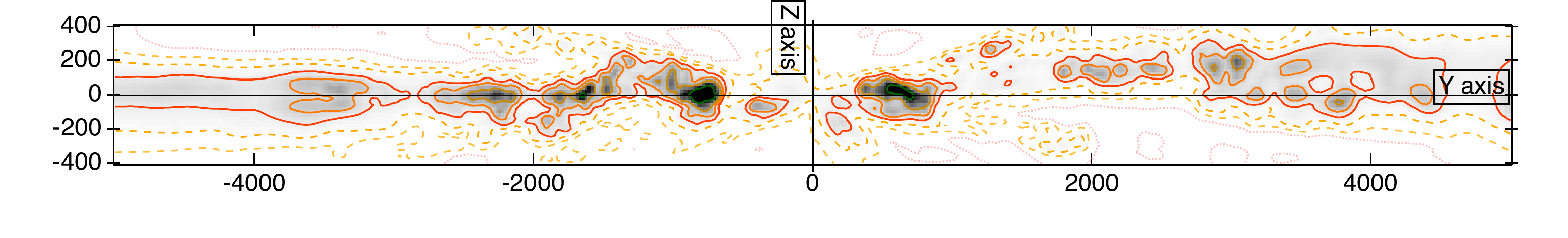}
    \caption{Inverted 3D dust distribution. {\bf Top:} Extinction density in the XY plane containing the Sun. The orientation is parallel to the Galactic Plane. The map is at the lowest resolution (correlation length of 50 pc). The Sun is at (0,0). Units are in parsecs. The colour scale is the same for all maps and is shown in Fig. \ref{plan25}. The thick dashed-dot black line shows the distance beyond which cells may be devoid of targets (see text for caveats at large distance). Iso-extinction density contours are drawn for $1 \cdot 10^{-5}$ (red), $5 \cdot 10^{-5}$ (dashed pink), $1 \cdot 10^{-4}$ (dashed light orange), $2 \cdot 10^{-4}$ (dashed orange), $5 \cdot 10^{-4}$ (red), $1 \cdot 10^{-3}$ (orange),  $2 \cdot 10^{-3}$ (brown), $5 \cdot 10^{-3}$ (dark green), $1 \cdot 10^{-2}$ (dashed light green), $2 \cdot 10^{-2}$ (blue) mag\, pc$^{-1}$. A giant cavity devoid of dust lies between longitudes l$\simeq$85\fdeg and l$\simeq$145\fdeg. {\bf Middle:} Extinction density in the XZ (meridian) plane. The Galactic centre is to the right. {\bf Bottom:} Extinction density in the YZ (rotation) plane. The sense of rotation is to the top.}
    \label{plan50}
\end{figure*}

\begin{itemize}
\item{10,000 pc x 10,000 pc x 800 pc for R= 50 pc}
\item{6,000 pc x 6,000 pc x 800 pc for R= 25 pc}
\item{3,000 pc x 3,000 pc x 800 pc for R= 10 pc,}
\end{itemize}  

where R is the size of the minimum correlation length, corresponding to the last iteration of the inversion. The extinction density maps can be explored and exploited using a dedicated application, called G-Tomo, deployed on the EXPLORE website \footnote{https://explore-platform.eu {\bf(use version 2 (v2) tools when available to obtain the distributions presented in this article}}. We illustrate the results by showing examples of extinction density images in three selected planes for the first, second, and third  map. The fourth map will be studied in more detail and compared with other tracers of the dense ISM in a forthcoming dedicated article. Fig.~\ref{plan50} displays the extinction density of the first map in the XY plane, parallel to the Galactic Plane and containing the Sun. The minimum correlation length of 50~pc allows showing only large structures. Several iso-contours of the extinction density are drawn to delineate the most opaque and most transparent areas. Because the number of targets is very limited beyond 3~kpc and because the distribution of these distant targets is very inhomogeneous, this map reveals structures only in several distant regions, while in others, the solution is strongly influenced by the homogeneous prior used in the first iteration of the hierarchical inversion. We drew an isocontour that delimits regions possessing a target density below one per 100$\times$100$\times$100~pc$^{-3}$ and where the inversion cannot be performed, even at the lowest spatial resolution. Fig.~\ref{plan50} also shows the extinction density from the same 3D map in vertical planes containing the Sun and oriented along the axis Sun-Galactic centre (the meridian plane) and along the tangential direction to the Sun circle (the rotation plane). We refer to \cite{Lall19,Lall22} for an identification of the main structures. In our former map based on Gaia e-DR3 and 2MASS photometric data \citep{Lall22}, we emphasised the wavy pattern of the dust distribution along the Z-axis. This pattern is seen in many regions. We also pointed out the similarity the mean vertical peak-to-peak amplitude (about 300 pc) and the vertical period of the spectacular snail-shaped stellar kinematical pattern discovered in Gaia data by \cite{Antoja18}. The more extended new map shows that the wavy pattern seems to disappear at large distances along the X-, -X-, and -Y-axes, and seems to persist along the +Y-axis (see Fig. \ref{plan50} middle and bottom). However, more data are needed to confirm the pattern at these distances.

\begin{figure*}
    \includegraphics[width=0.95\linewidth]{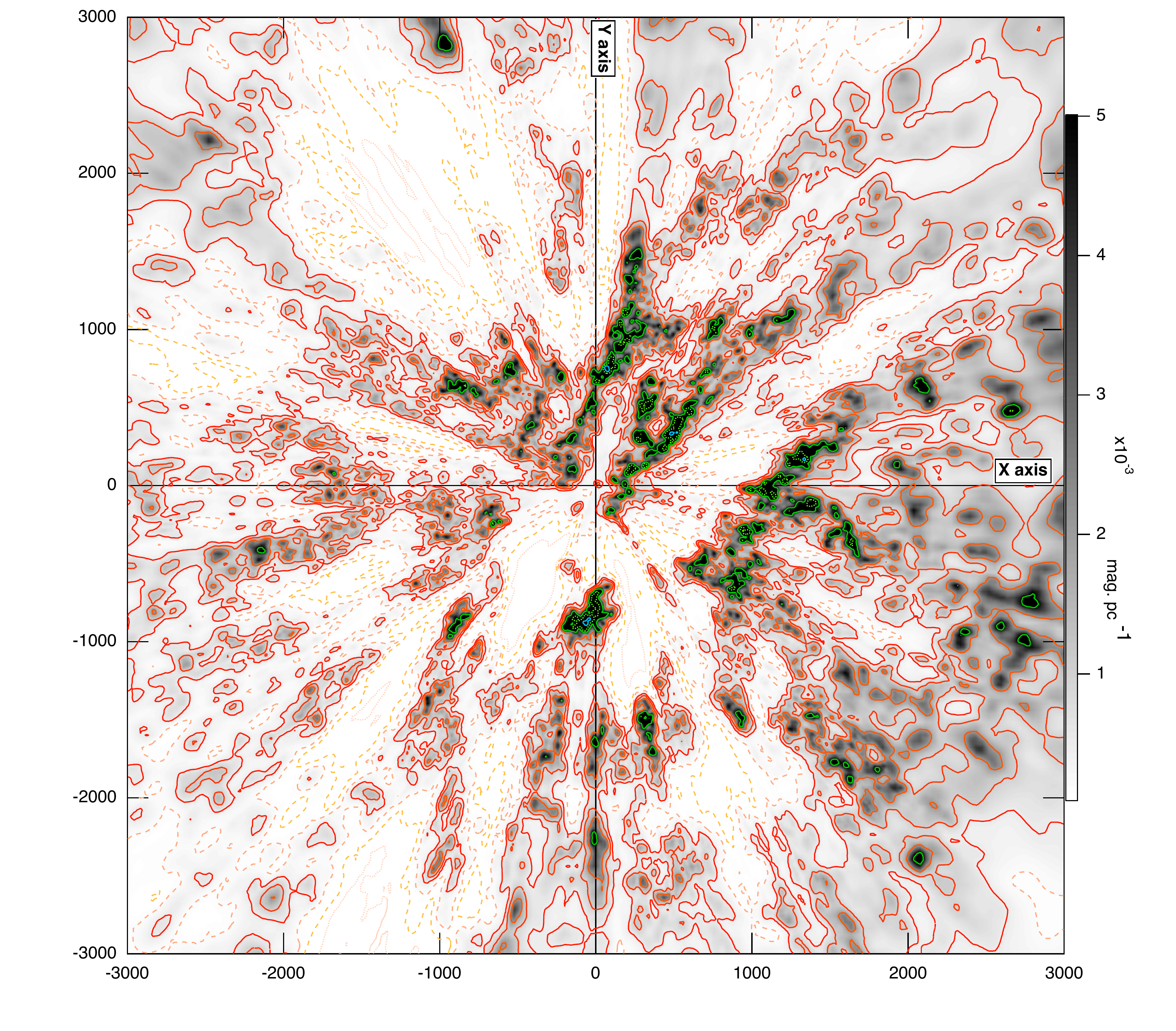}
    \centering
    \includegraphics[width=0.98\linewidth]{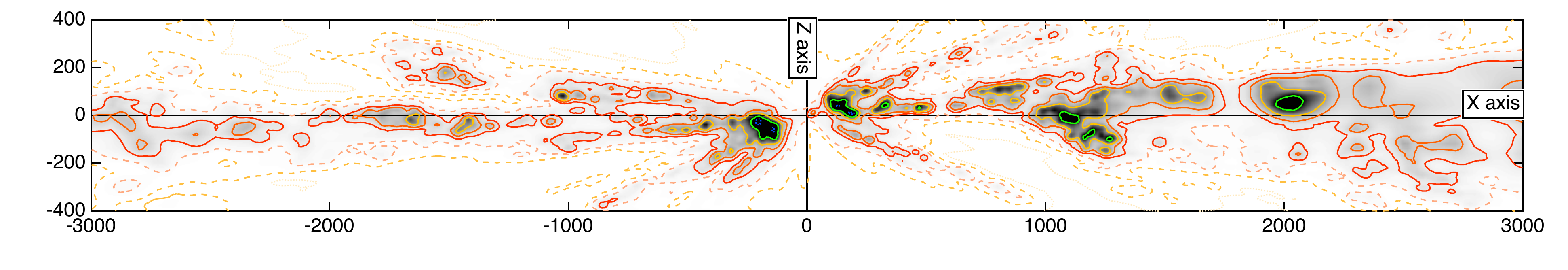}
    \includegraphics[width=0.98\linewidth]{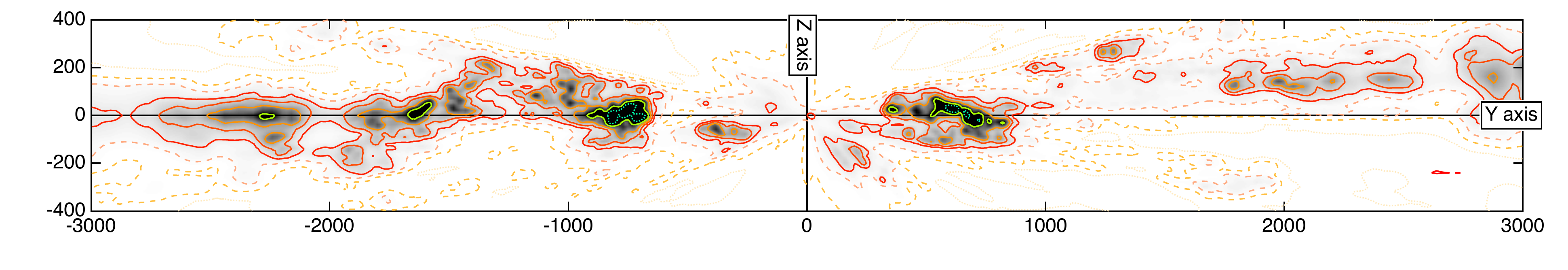}
    \caption{Same as Fig. \ref{plan50} for the less extended 6kpc x 6kpc x 0.8 kpc map inverted with a correlation length of 25~pc. The colour scale is the same for all images.}
    \label{plan25}
\end{figure*}

Fig.~\ref{plan25} is identical to Fig.~\ref{plan50} for the second map, which covers only 6~kpc by 6~kpc along the Plane. Here the maximum resolution, corresponding to a correlation length of 25~pc, allows showing many more details. As explained above, the maximum resolution is not achieved everywhere. In the frame of the hierarchical technique, the resolution is even limited by the target density. For this reason, the highest resolution is achieved in regions with high target density, and the map is more detailed than the previous one. Conversely, in regions with very few targets, the map is similar to the previous one.

In addition to structures discussed in previous works, the two maps in Fig. \ref{plan25} and Fig. \ref{plan50} reveal a gigantic, oval cavity in the second quadrant very clearly. This 2.5~kpc by 2.0~kpc region devoid of dust has previously been overlooked because its far boundary was not clearly delineated because the mapping extent was limited or the resolution was too poor. Here its entire contours are drawn rather precisely. This remarkable, huge structure is oriented very differently from other cavities and chains of clouds, and it suggests a local phenomenon, at variance with structures triggered by external perturbations (dwarf galaxy crossing, bar and arm resonances; see \cite{Antoja18, Khoperskov20}. The most likely origin for this huge cavity is a large series of supernovae. Comparisons with O, B star associations will help tracing the history of this region. 

\begin{figure*}
    \includegraphics[width=0.95\linewidth]{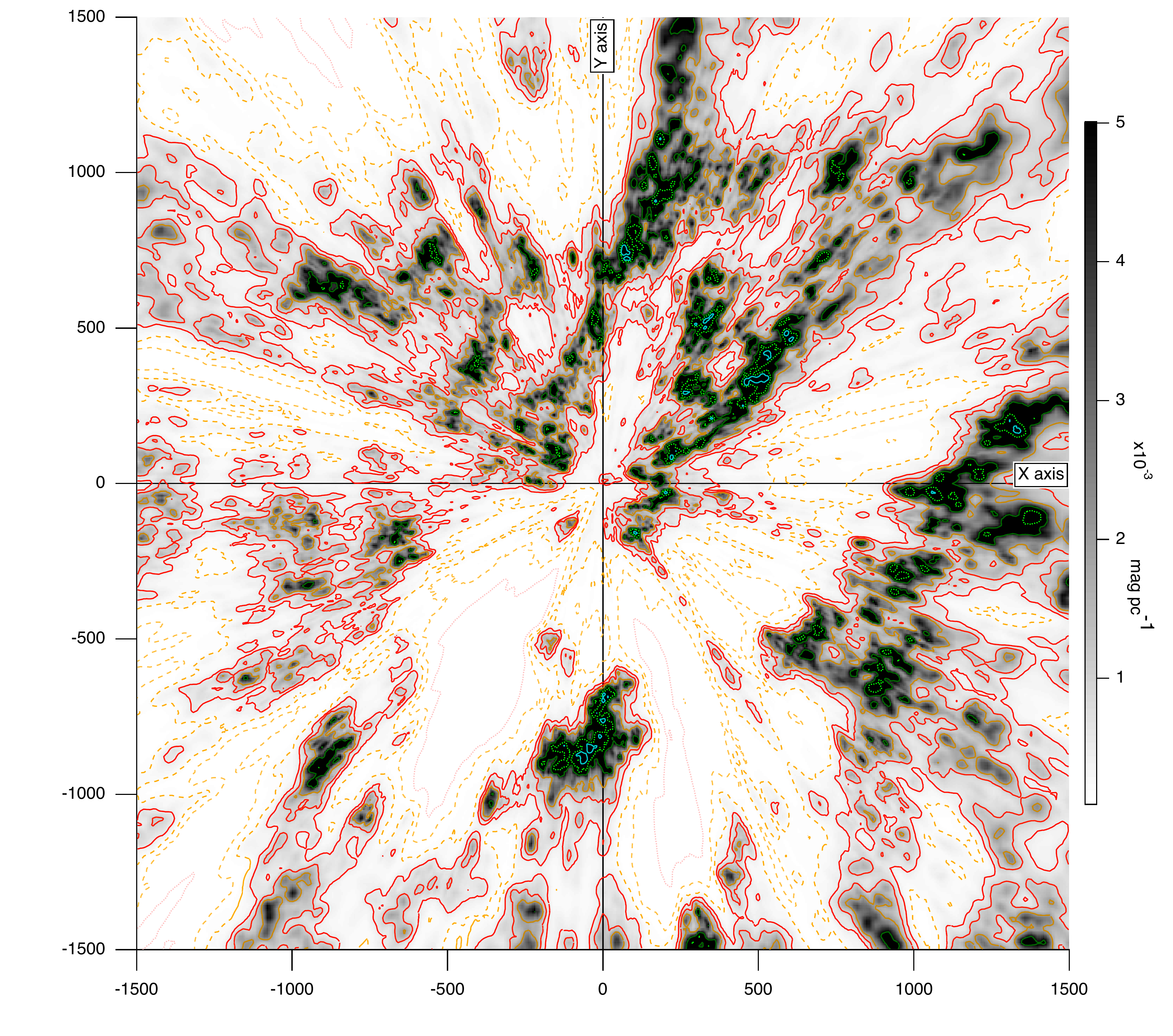}
    \includegraphics[width=0.95\linewidth]{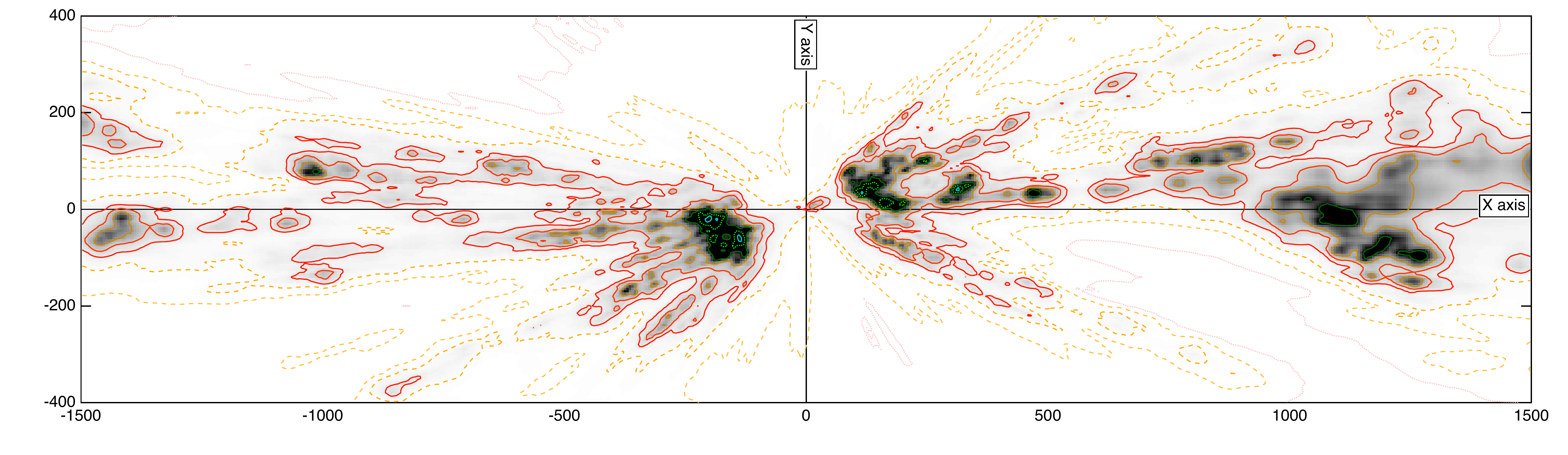}
    \includegraphics[width=0.95\linewidth]{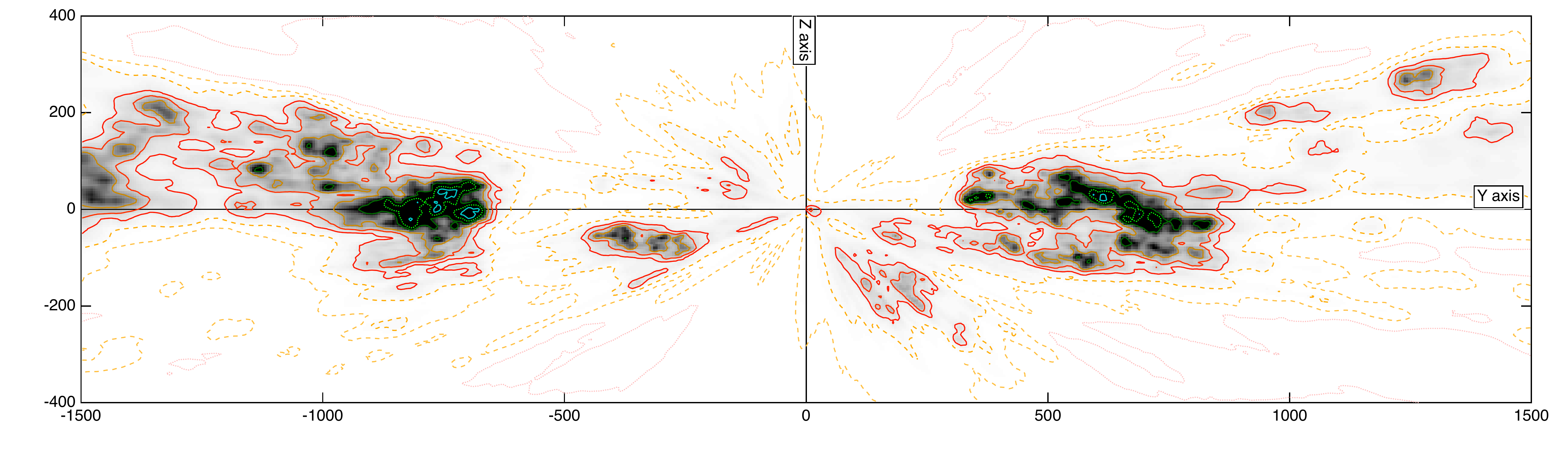}
    \caption{Same as Fig.~\ref{plan25} for the 3kpc x 3kpc x 0.8 kpc map inverted with a correlation length of 10~pc.}
    \label{plan10}
\end{figure*}

Fig.~\ref{plan10} is similar to Fig.~\ref{plan50} and Fig.~\ref{plan25} for the third map. It covers only 3~kpc by 3~kpc and shows the result of a hierarchical inversion with a smaller final correlation length of 10 pc. 

Isocontours of equal extinction densities are drawn on each map for the same series of extinction density levels (see Fig.~\ref{plan50} for details). Contours delineating the densest regions appear gradually from the lower to the higher resolution map. This is a natural consequence of the decreasing correlation kernel, and it arises because the integrated extinction can be distributed in smaller volumes in the high-resolution maps.

\section{Summary and discussion}\label{discussion}

We have devised a technique of inter-calibrating extinction catalogues that were based on different data and different techniques of extinction estimates. Our goal was suppressing biases or artefacts in 3D maps of extinction density that are produced by the inversion of merged catalogues. The inter-calibration was based on extinction comparisons  for targets from different catalogues that are located in the same region of 3D space. It was based on the principle that all estimated extinctions should be equal at the same location, regardless of the method, input data type, and stellar type. The comparison was conducted cell by cell in a subspace constituted by the two first components of a PCA initially performed in the (G, G$_{B}$, G$_{R}$, J, H, and K) multi-colour space. One main advantage of this technique is its potential application to two catalogues for which the targets do not overlap. This allows using catalogues based on photometry in different wavelength ranges. The only requirement is that in at least one region of space targets from the different sources co-exist and are numerous enough to establish some statistical relations. Outside this region, other targets may be distributed differently. This is a second advantage because it allows the use of catalogues that consist of targets that are mostly located in different regions of space.

As an illustration, we used as the catalogue of reference for the inter-calibration a spectrophotometric catalogue, and we performed the inter-calibration on a purely photometric catalogue, previously presented in \cite{Lall22}. The combination of spectroscopy and photometry provides a more accurate extinction, and the inter-calibration allows transferring part of this accuracy to the other data. In order to reduce the dimension of the problem, a PCA was performed in the (G, G$_{B}$, G$_{R}$, J, H, and K) space by associating colours used for the constitution of the catalogue. The subspace constituted by the two first components was split into cells, in which deviations from the reference were estimated. These deviations were computed using all targets from the reference catalogue located at a short distance of each secondary target in 3D space. Corrections and filtering were deduced in each cell in multi-colour space. The same technique can be applied to a different multi-colour space or to any multi-dimensional space associating the different types of measurements entering the determination of the extinction.

We have used several validation tests and showed that the technique produces a merged catalogue that is significantly more self-coherent by comparison with a simple concatenation of data, including in regions of space that are devoid of reference targets. We emphasise that the same method can be also successfully applied to a single dataset to detect biases and increase the internal self-coherence. We also note that this work contains some arbitrary choices of reference values. Further work is planned on the various criteria and the search for the best references. Gaia Data Release 3 will be used for this purpose.

We have inverted the merged catalogue to produce several 3D distributions of extinction density in various volumes and for different resolutions comprised between 50 pc and 5 pc, following the technique presented in \citep{Vergely10} and \cite{Lall19}. The addition of spectroscopic data allows us to reach larger distances from the Sun and to improve the quality of the map. The wavy pattern around the Z-axis is confirmed and better delimited. Based on our previous maps \citep{Lall22}, we suggested that it is connected with the spectacular snail-shaped stellar kinematical pattern discovered in Gaia data \citep{Antoja18}, whose origin is currently debated. Future Gaia data and further studies are expected to shed light on this peculiar local pattern. Some cavities are more clearly revealed, in particular, a 2.5 kpc by 2 kpc region that is fully devoid of dust in the second quadrant. Further work is needed to understand its oval shape, which is more suggestive of a superbubble than of an inter-arm region. New Gaia data for the locations of massive stars and 3D motions will help trace the formation of this huge cavity.

\begin{acknowledgements}

We thank our referee for his very careful reading of the manuscript and his many constructive remarks which helped to improve the clarity of the text and the presentation of the figures.\\
J-L.V. and N.L.J.C. acknowledge support from the EXPLORE project. EXPLORE has received funding from the European Union’s Horizon 2020 research and innovation programme under grant agreement No 101004214.\\
J-L.V. acknowledges support from the THETA Observatoire des Sciences de l'Univers in Besan\c{c}on.\\
This research or product makes use of public auxiliary data provided by ESA/Gaia/DPAC/CU5 and prepared by Carine Babusiaux.\\
This research has made use of the SIMBAD database, operated at CDS, Strasbourg, France.\\
This work has made use of data from the European Space Agency (ESA) mission Gaia (https://www.cosmos.esa.int/gaia), processed by the Gaia Data Processing and Analysis Consortium \\
(DPAC, https://www.cosmos.esa.int/web/gaia/dpac/consortium). Funding for the DPAC has been provided by national institutions, in particular the institutions participating in the Gaia Multilateral
Agreement. This work also makes use of
data products from the 2MASS, which is a joint project of
the University of Massachusetts and the Infrared Processing
and Analysis Center/California Institute of Technology,
funded by the National Aeronautics and Space Administration
and the National Science Foundation.\\
\end{acknowledgements}

\bibliographystyle{aa} 
\bibliography{mybib}

\appendix

\section{Extinction bias estimate from absorption lines in white dwarf UV spectra}\label{appendix:A}

\begin{figure}
\includegraphics[width=0.99\linewidth]{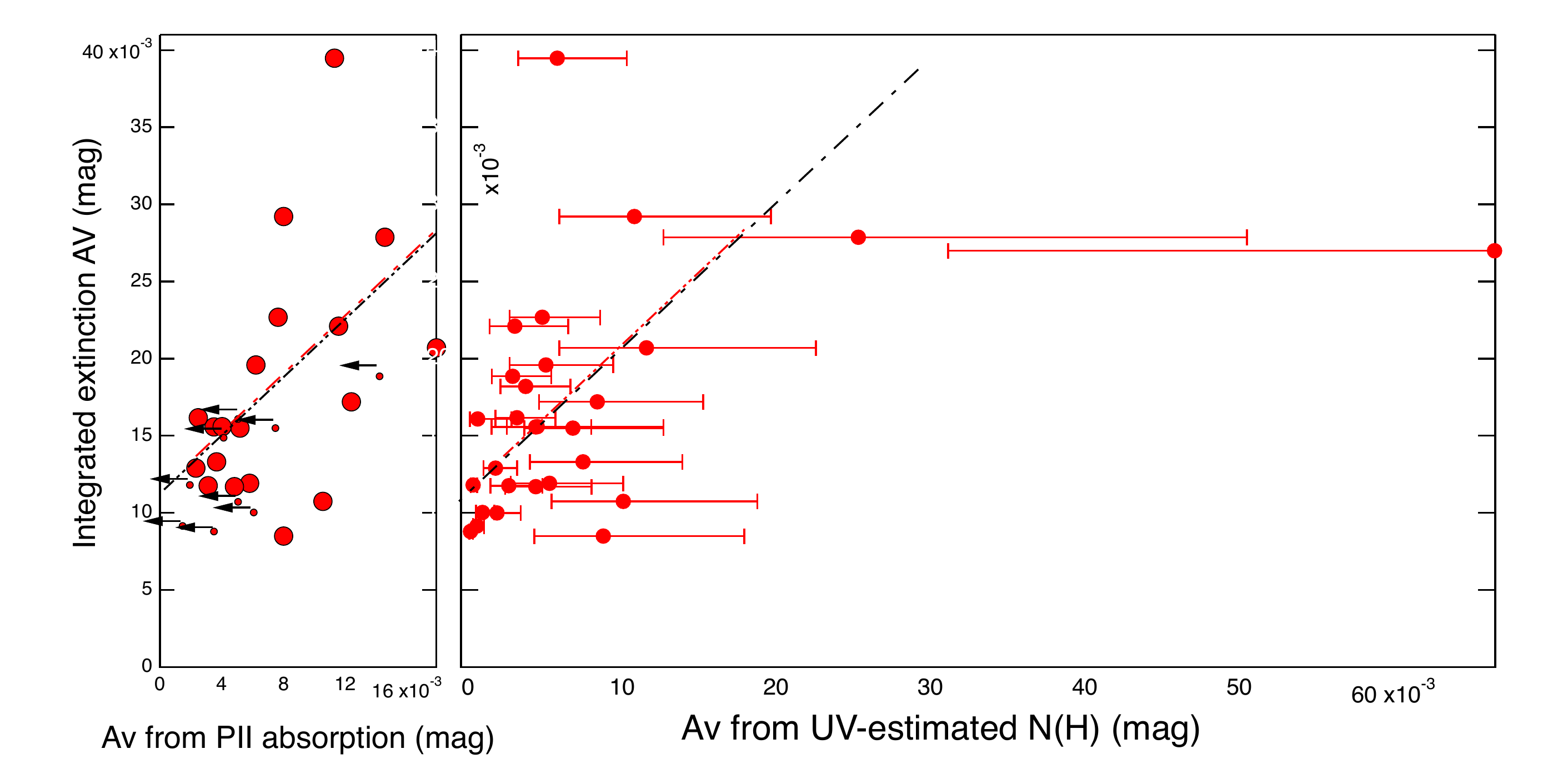}
    \caption{Comparison between extinctions estimated from UV absorption lines in the spectra of nearby white dwarfs and extinction values integrated in a preliminary 3D map (see text). {\bf Left:} Using \ion{P}{ii} ion columns from \cite{Lehner03} and phosphorus depletions from \cite{Jenkins09}. Black arrows correspond to estimated negative values and show the corresponding upper limits. The dot-dashed black line is a linear fit using only positive values. {\bf Right:} Using N(H) columns estimated by \cite{Jenkins09} from P, O, Si, and Fe ions. The dot-dashed black line is reported from the left panel. Both estimates indicate a positive bias on A$_{V}$  of about 0.01~mag.}
    \label{biasfromwds}
\end{figure}

Extinctions estimated by \cite{Sanders18} are forced to be positive, which may induce a small bias at very low extinctions. Because this may introduce systematic effects during the combination with the \cite{Queiroz20} catalogue, in which extinctions are allowed to be negative, we searched for a quantification of a potential small bias. To do this, we performed a preliminary inversion of the \cite{Sanders18} catalogue alone, using the 3D distribution from GAIA-2MASS photometry \citep{Lall19} as a prior. Because the error bars of the \cite{Sanders18} catalogue are very small, its content constrains the 3D distribution of extinction at very short distance, where the prior has no effect. As a result, if a bias does exist, it should be reflected in the opacity density at the short distances found in this preliminary inversion. 

In parallel, we searched for gaseous absorption lines in the UV spectra of nearby WDs with the goal of estimating the extinction at short distance in a different way, assuming a classical relation between dust and gas. An appropriate species is the \ion{P}{ii} ion, which has a small depletion and has been shown to be relatively well correlated with the gas column \citep[see ][and references therein]{Jenkins09}. We used the results from \cite{Lehner03}, and in particular, measurements of \ion{P}{ii} ion columns as well as the phosphorus depletion estimates from \cite{Jenkins09} to transform the \ion{P}{ii} columns into N(H), then into extinction A$_V$ using A$_V$ = 3.1* N(H) / $6 \times 10^{21}$ for about 30 nearby WDs. Fig.~\ref{biasfromwds} (left) shows the extinction integrated in the 3D map described above between the Sun and each WD as a function of the extinction estimated from \ion{P}{ii} absorbing columns along the path to the targets. Although the uncertainties on both quantities are large, the extrapolation of the average relation between the two quantities shows that for null columns of gas, the extinction integrated in the map is positive and about 0.01~mag, suggesting a bias on this order. 

Later on, and for most of the \cite{Lehner03} nearby WDs, \cite{Jenkins09} estimated the gas column using all available absorption data for P, O, Si, and Fe in combination. Fig.~\ref{biasfromwds} (right) shows the extinction integrated towards the slightly different set of WDs as a function of the extinction from N(H) using the same A$_{V}$/N(H) relation as above. The bias deduced in this way is similar. As a result, we removed 0.01~mag from all the extinction measurements from the \cite{Sanders18} catalogue.

\section{Merging the two extinction catalogues based on spectroscopy and photometry}\label{appendix:B}

As described in section \ref{spectrodata}, the methods used by \cite{Sanders18} and \cite{Queiroz20} to estimate the stellar parameters and extinctions differ in several aspects. Due to the use of a prior on the 3D distribution of extinction in the former catalogue and the existence of a global quality flag ({\it best}), we kept all good-quality data ({\it best} = 1) and removed the small bias described in the previous section. In the case of the latter catalogue, we performed several selections and searches for biases, as described below. 


\subsection{Data selection}
In the case of the \cite{Queiroz20} catalogue, we made use of the various flags tabulated by the authors and additionally excluded some targets based on the Gaia measurements. We list below the criteria for target exclusion.
\begin{itemize}
    \item For APOGEE data, the absence of the {\it Avprior} input.
    \item For all surveys, the presence of {\it uncal}, or {\it TEFF-LOGG-m-h-ALPHA-PARALLAX-Ggaia,} in the input flags.
    \item For all surveys, the presence of {\it bad}, {\it high,}  or {\it warn} in the output flags.
\item When Gaia parallaxes are used,
\begin{itemize}
    \item {\it ruwe} $\geq$ 1.4, or  {\it ipd frac multi peak} $\geq$2
\item a high bp-rp flux excess factor, namely C $\geq 3 \sigma(G) $ with C and $\sigma(G)$ computed following Table 2 and Equ. 18 from \cite{Riello21}
\end{itemize}
\item The presence of bright, angularly close targets being a potential source of bias during the spectroscopic observation. We used Gaia eDR3 astrometry and photometry to select stars within 10 $\mu$arcsec from each catalogue target. We computed a {\it weight} function $W$ of these secondary stars in the form of a sum of their potential contaminating flux $Cf$. We used for Cf the G-band flux fG and the angle $\theta$ with the catalogue target, \\
$W= \Sigma fG_{i} / fG^{0} /\theta_{i} $\\
where fG$^{0}$ is the G-band flux of the catalogue target. We excluded targets with $w\geq$0.7. This quantity was adjusted after a search for visible effects of a high value of $w$ in the form of outliers in line-of-sight extinction profiles. Based on this criterion, we excluded about 7,500 targets.
\end{itemize}

\subsection{Additional empirical corrections using the dust emission}

\begin{figure}
   \includegraphics[width=0.99\linewidth]{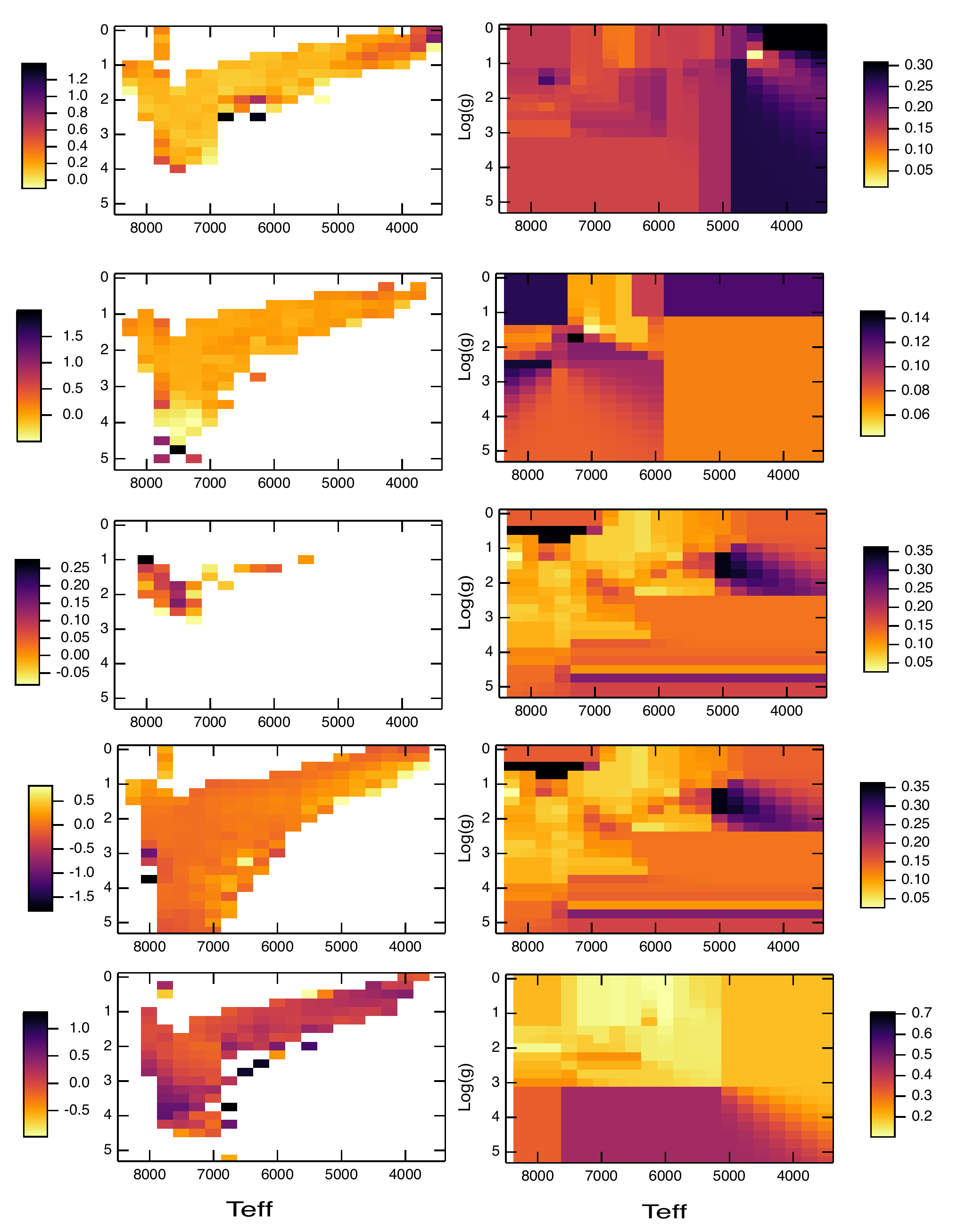}
   \caption{Empirical corrections of the extinction (in mag) and additional uncertainties based on low dust emission regions (see text). {\it Left:} From top to bottom:  APOGEE, GALAH, GES, LAMOST, and RAVE empirical corrections. Blank cells correspond to an absence of targets. No correction was made for other targets with higher dust emission that lay in the same cell. The colour scale varies from one survey to the other. {\it Right:} Additional empirical uncertainties. Because it contains very few targets, we arbitrarily added uncertainties similar to those of LAMOST for the GES survey.}
    \label{correcplanck}
\end{figure}

We used the dust optical depth at 353 GHz measured by Planck \citep{PlanckXI} to select directions of very weak extinction. The selection criterion was $\tau(353)\leq 2.1$ $10^{-6}$, which corresponds approximately to A$_V$ $\leq$ 0.1~mag, according to the average relation E(B-V) = $1.5 \cdot 10^{4}$ $\tau(353)$ from \cite{Remy18} and assuming A$_V$ = 3.1 E(B-V). All the selected sight lines correspond to high latitudes, and for this reason, the large majority of the target stars are located beyond the dust layer. For all these sightlines, we removed from the catalogue extinction A$_V$(c) the weak amount of extinction corresponding to the Planck optical depth A$_V$(P), using the above relation. For each catalogue separately, we then computed the average difference A$_V$(c)-A$_V$(P) for targets belonging to temperature-gravity log($g$)-T$_\mathrm{eff}$ two-dimensional cells. The T$_\mathrm{eff}$ grid extends from 3500 to 9000~K in steps of 250~K, and the log($g$) grid extends from 0 to 5 by steps of 0.25. The  distribution of A$_V$(c)-A$_V$(P) in each cell was fitted to a Gaussian distribution. The central value was considered as a bias for the corresponding cell in the corresponding survey. The standard deviation was added quadratically to the catalogue uncertainty for all stars in the T$_\mathrm{eff}$-log($g$) cell (regardless of the Planck value). Fig.~\ref{correcplanck} shows the values of the biases determined using this empirical method. Some cells are empty and correspond to numbers of targets that are too low to allow a bias estimate. 

A second filtering and correction was performed on the basis of the Gaia photometric fluxes. To do so, we used the updated Gaia eDR3 values. After the above correction we distributed all targets in directions of low dust emission in G$_{B}$-(G$_{B}$-G$_{R}$) two-dimensional cells. Here again, one would expect a distribution of  A$_V$(c)-A$_V$(P) centred around zero. We fitted the distribution to a Gaussian and estimated a bias and an additional uncertainty in the same way than for the log($g$)-T$_\mathrm{eff}$ bins. Additionally, for each interval for G$_{B}$, we excluded targets in (G$_{B}$-G$_{R}$) cells with very high bias or dispersion. This last step was performed by eye. The intervals, biases, additional uncertainties, and exclusions are  listed in Table~\ref{corrfrommag}. 

\begin{figure}
    \includegraphics[width=0.99\linewidth]{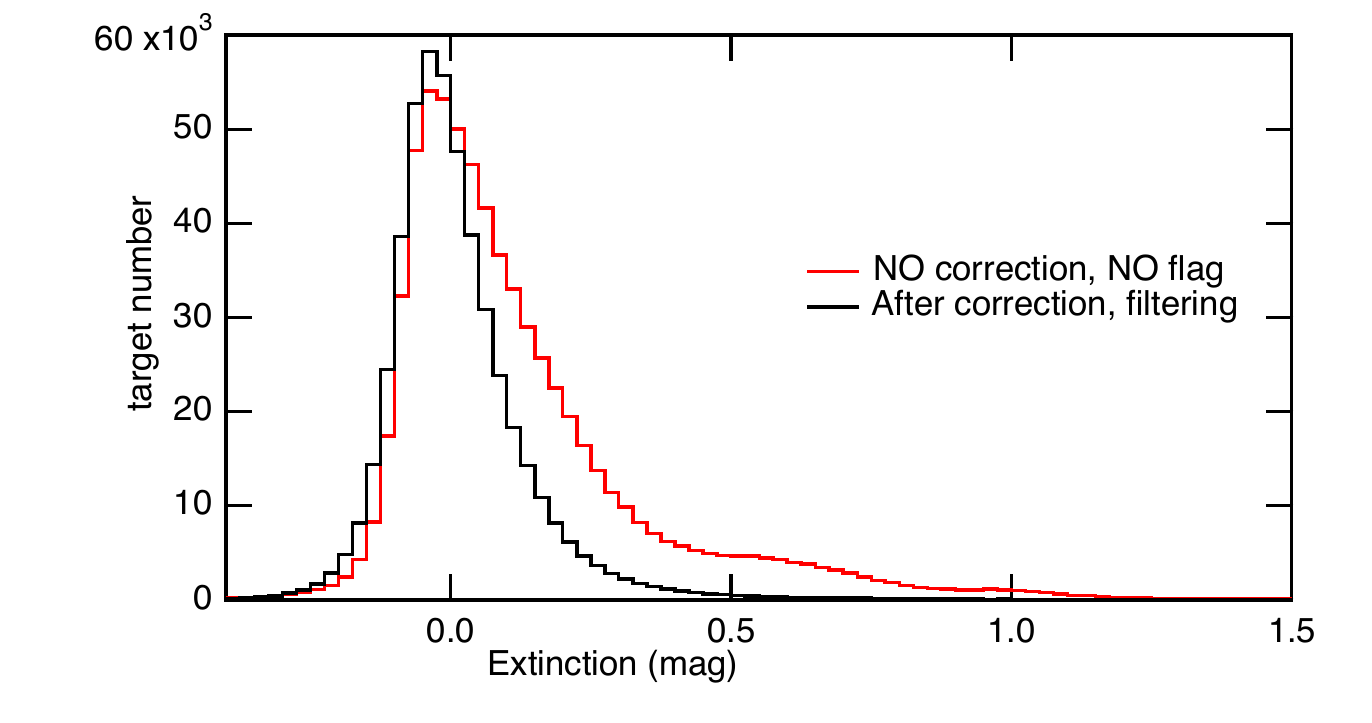}
    \caption{Histogram of the extinctions for the set of stars with low dust optical thickness before and after filtering and introduction of empirical corrections.}
    \label{improve}
\end{figure}

Fig \ref{improve} shows the histogram of all extinctions A$_V$(c)-A$_V$(P) before and after application of the filtering and empirical corrections for all stars with $\tau(353)\leq 2.1$ $10^{-6}$. About 25~\% of the targets are excluded if all constraints are applied. The wings of the distribution were largely reduced and the width of the distribution has decreased.

\begin{table*}
\caption{Second empirical correction based on low-extinction regions: For each G$_{B}$ magnitude interval, we list the allowed interval for G$_{B}$-G$_{R}$, the bias, and the standard deviation to be quadratically added to the other terms.}
\begin{tabular}{cccccccccccc}
\hline
G$_{B}$ interval & 9-10 & 10-11 & 11-12 & 12-13 & 13-14 & 14-15 & 15-16 & 16-17 & 17-18 & 18-19 & 19-20\\
G$_{B}$-G$_{R}$ & >0.58 & 0.5-1.5 & >-0.09 & 0.02-1.63 & 0.02-1.63 & 0.03-1.54 & 0.03-1.45 & 0.0-1.5 & 0-1.55 & 0.6-1.5 & <1.1 \\
Corr-Av & 0.0329 &0.0324& 0.0513 & 0.0547& 0.0504 & 0.0418& 0.0316& 0.0273&0.0088 & -0.0244& -0.1044\\
$\sigma$(c)& 0.09 & 0.10 & 0.09 & 0.08 &0.08 & 0.09 & 0.105 & 0.115 & 0.13 & 0.19 & 0.34
\label{corrfrommag}
\end{tabular}
\end{table*}

\end{document}